\documentclass[notoc]{JHEP3}
\usepackage{graphicx}
\usepackage{color,cite}

\newcommand{\nbar}{ {\bar{n}}}

\newcommand{\ds}{\displaystyle}

\newcommand{\nn}{\nonumber \\}

\newcommand{\eqref}[1]{(\ref{#1})}
\newcommand{\text}[1]{{\rm #1}}

\newcommand{\bS}{\hat{S}}
\newcommand{\rS}{S_r}
\newcommand{\GNGL}{\Gamma^{\mathrm{NGL}}}
\newcommand{\gsreg}{\gamma^{\mathrm{NGL}}}

\newcommand{\thR}{R}
\newenvironment{changemargin}[2]{%
  \begin{list}{}{%
    \setlength{\topsep}{0pt}%
    \setlength{\leftmargin}{#1}%
    \setlength{\rightmargin}{#2}%
    \setlength{\listparindent}{\parindent}%
    \setlength{\itemindent}{\parindent}%
    \setlength{\parsep}{\parskip}%
  }%
  \item[]}{\end{list}}

\def\eqn#1{Eq.~(\ref{eq:#1})}

\def\eqns#1#2{Eqs.~(\ref{eq:#1}) and~(\ref{eq:#2})}

\def\fig#1{Figure~{\ref{fig:#1}}}
\renewcommand{\sec}[1]{Section~\ref{sec:#1}}
\newcommand{\ssec}[1]{Section~\ref{ssec:#1}}
\newcommand{\be}{\begin{equation}}
\newcommand{\ee}{\end{equation}}
\newcommand{\cmb}{\begin{changemargin}}
\newcommand{\cme}{\end{changemargin}}
\newcommand{\bea}{\begin{eqnarray}}
\newcommand{\eea}{\end{eqnarray}}

\newcommand{\ecm}{E_{\rm CM}}

\def\spa#1.#2{\langle#1\,#2\rangle}
\def\spb#1.#2{[#1\,#2]}
\def\sandmm#1.#2.#3{%
\left\langle\smash{#1}{\rphantom1}\right|{#2}%
\left|\smash{#3}{\rphantom1}\right]}
\def\spab#1.#2.#3{\sandmm#1.#2.#3}
\def\spba#1.#2.#3{\sandpp#1.#2.#3}
\def\spaa#1.#2.#3.#4{\sandmp#1.{#2#3}.#4}
\def\spbb#1.#2.#3.#4{\sandpm#1.{#2#3}.#4}
\def\spash#1.#2{\spa{\smash{#1}}.{\smash{#2}}}
\def\spbsh#1.#2{\spb{\smash{#1}}.{\smash{#2}}}

\def\ksl{\not{\hbox{\kern-2.3pt $k$}}}
\def\lsim{\lesssim}

\def\in{{\rm in}}
\def\out{{\rm out}}

\newcommand{\rd}{\mathrm{d}}

\preprint{IFT-UAM/CSIC-11-96}
\title{Resummation of Jet Mass at Hadron Colliders}

\author{Yang-Ting Chien, Randall Kelley and Matthew D. Schwartz\\
Center for the Fundamental Laws of Nature,\\
Harvard University,\\
Cambridge, MA 02138, USA}
\author{Hua Xing Zhu\\
Department of Physics and State Key Laboratory of Nuclear Physics and Technology,\\
Peking University,\\
Beijing 100871, China}

\abstract{
A method is developed for calculating the jet mass distribution at hadron
colliders using an expansion about the kinematic threshold. In particular, we
consider the mass distribution of jets of size $R$ produced in association
with a hard photon at the Large Hadron Collider. Expanding around the kinematic
threshold, where all the energy goes into the jet and the photon, provides
a clean factorization formula and allows for the resummation of logarithms
associated with soft and collinear divergences. All of the large logarithms
of jet mass are resummed at next-to-leading logarithmic level (NLL), and all
the global logarithms at next-to-next-to-leading logarithmic level (NNLL). A
key step in the derivation is the factorization of the soft function into
pieces associated with single scales and a
remainder which contains non-global structure. This step, which is standard
in traditional resummation, is implemented in effective field theory which is
then used to resum the large logarithms using the renormalization group in a
systematically improvable manner.
}

\begin{document}
\section{Introduction}
\label{sec:intro}
The properties of jets produced at the Large Hadron Collider (LHC) have
been coming into increased focus over the last few years, both from
theoretical and experimental points of view. Jet properties can be used
both to understand QCD and to find and discriminate among models of
new physics. Indeed, many methods involving jet substructure have been
developed over the last few years that can be used in new physics searches
\cite{Butterworth:2002tt,Butterworth:2008iy,Kaplan:2008ie,Ellis:2009me,Thaler:2008ju,Krohn:2009th,Gallicchio:2010dq,Thaler:2010tr,Gallicchio:2010sw,Cui:2010km,Gallicchio:2011xq,Altheimer:2012mn,Ellis:2012sn}.
These studies are almost all based on Monte-Carlo
simulations. As the precision of the experimental measurements improves,
precision calculations beyond the level of current Monte-Carlo generators
will become increasingly important. In this paper we provide a framework in
which the simplest jet property, jet mass, can be computed in a systematically
improvable way.

Precision QCD calculations of jet shapes, such as jet mass, are difficult for
a number of reasons. Calculations at fixed order in perturbation theory are
only useful at large values of jet mass in the tail of the distribution. Near
the peak, large Sudakov double logarithms become increasingly important. The
resummation of these logarithms is critical in achieving distributions with
even qualitative agreement with data. However, the resummation of jet mass
in a realistic hadronic collider environment is challenging due to the large
number of variables: the jet size $R$, the jet algorithm, the beam remnants,
the initial state radiation, hadronization, underlying event, etc.

There has been flurry of work on jet mass resummation over the last few years
\cite{Becher:2008cf,Cheung:2009sg,Ellis:2009wj,Jouttenus:2009ns,Ellis:2010rwa,Banfi:2010pa,Chien:2010kc,
Kelley:2011tj,Jouttenus:2011wh,Li:2011hy,KhelifaKerfa:2011zu,Kelley:2011aa,
Li:2012bw}. In particular, it has been demonstrated that
large logarithms of jet masses can be understood and in some cases resummed
using effective field theory techniques. In this paper we consider the simplest
QCD event shape, the jet invariant mass, and we focus on a very simple event
topology: a jet produced in association with a hard photon. This process
provides a natural setting to generalize similar work that has been done for
$e^+ e^-$ colliders. Furthermore, such a simple final state allows us to
explore issues such as dealing with the beam remnant and the jet size, before
attempting to extend these results to more complicated multi-jet final states.

Many of the ingredients necessary for jet mass resummation in direct
photon production have already been provided in previous work that
calculated the direct photon $p_T$ and rapidity distribution using threshold
resummation~\cite{Becher:2009th}. For the $p_T$ spectrum inclusive over the jet
properties, one can treat the jet as everything-but-the-photon.
Since the beam remnants are included in the jet, the factorization formula in
the inclusive case is particularly simple. In this paper, we want to calculate
a cross section which is differential in the jet mass. Thus, we must separate
the soft radiation into an in-jet region and an out-of-jet region,
which leads to a modified factorization
formula.

Although the jet mass distribution in this paper is calculated by expanding
around the machine threshold limit, where the jet and photon have the maximum
possible energy, the results will be accurate well away from threshold, at
phenomenologically relevant values of the jet momentum, due to dynamical
threshold enhancement~\cite{Appell:1988ie,Catani:1998tm,Becher:2007ty}. Dynamical threshold enhancement works because of the
extremely rapid fall-off of the parton distribution functions (PDFs) as momentum
fractions approach unity. This fall off forces radiation in realistic
events to have similar properties to radiation in events near
threshold. In particular, the large logarithms associated with soft and
collinear singularities are determined by an effective maximum energy, set by
a parton threshold, rather than the hadronic threshold. The power of threshold
resummation has been confirmed by direct comparison with data in many cases.
Here, in the absence of direct photon jet mass data, we confirm it by comparing
to {\sc pythia} \cite{Sjostrand:2006za}.

When dealing with an exclusive observable such as the jet mass, one is forced
to contend with multiple relevant scales: the hard scale of the partonic
collision, the jet scale associated with collinear radiation in the jets,
the factorization scale associated with collinear radiation in the beams,
and various soft scales. Through the use of the soft-collinear effective
field theory (SCET) \cite{Bauer:2000yr,Bauer:2001yt}, each of these scales is
associated with a function that accounts for the physics of each scale. The
multiple soft scales are particularly troublesome to deal with because the
factorization formula coming out of effective theory does not distinguish them:
SCET gives you only one soft function, with multiple scales. Previous work on
multi-scale soft functions in SCET has observed that, in the absence of an
exact two loop soft function calculation, improved agreement with QCD can result
from assuming a factorized soft function \cite{Kelley:2011tj,Kelley:2011aa}. In
this paper, we make further progress in understanding this refactorization by
taking inspiration from results on resummation in perturbative QCD.

Previous work on jet mass resummation in perturbative QCD, such as in
\cite{Almeida:2008tp,Li:2011hy,Li:2012bw} has approximated the jet as
having only collinear radiation and soft-collinear radiation. These two
ingredients combine into an infrared finite object that is a type of jet
function, but different from the jet function used in SCET. This simple
approximation leads to good agreement with CDF data in~\cite{cdfjetmass} and
indicates that the dominant large logarithmic corrections are contained in
the collinear and soft-collinear sectors. To go beyond the leading-logarithmic
accuracy of~\cite{Li:2011hy,Li:2012bw}, we need to also include initial
state radiation and color coherence effects. These are naturally part of the
threshold calculation we provide here. We therefore factorize out of the
full soft function a part associated with radiation which goes entirely into the jet, which
has its own natural scale. We give a precise operator definition of this
regional soft-function that, in particular, contains all the modes
that are simultaneously soft and collinear to the jet.
Thus our results will reduce to previous results
at leading-logarithmic level, but are systematically improvable.

Our systematic improvements over previous work include the full
next-to-next-to-leading logarithmic (NNLL) resummation of the global logarithms,
including initial state radiation and color-coherence effects.
One difficult unsolved problem in jet observables which are not
completely inclusive is that of non-global logarithms (NGLs)
\cite{Dasgupta:2001sh,Dasgupta:2002bw,Banfi:2002hw,Appleby:2002ke}. Although we
have not succeeded in resuming NGLs in this work, and thus our
distribution is only formally valid at the NLL level, our factorized
soft function clarifies the role of the missing terms. In an related observable, jet
thrust~\cite{Kelley:2011tj,Kelley:2011aa}, the anomalous dimension of the analogy of our regional soft function
is proportional to the leading non-global logarithm. We use the correspondence
to estimate the size of NGLs for jet mass by varying the appropriate
coefficient in our expressions. The effect of non-global logarithms is
significant in the peak region and must be understood better for further systematic improvements.

Because the factorization formula in SCET gives different objects with different associated
scales, we can resum large logarithms by matching and running between these
scales. It is an advantage of the effective theory approach that these scales are manifest throughout the
calculation.
Due to the convolution with the non-perturbative PDFs, scale choices
cannot be made analytically. We therefore choose our scales numerically,
following the approach of~\cite{Becher:2007ty}.
As emphasized in~\cite{Becher:2011fc} and \cite{Becher:2012xr}, it
is an advantage of the effective theory approach that natural scales automatically appear through this
numerical procedure, in contrast to fixed order calculations where they can only be guessed.
Our results are compared to the output of {\sc pythia}~\cite{Sjostrand:2007gs}, with very good agreement.

\section{Kinematics and the Observable}
\label{sec:kinematics}

\begin{figure}[t]
    \begin{center}
    \includegraphics[scale=0.6, trim = 0mm 14mm 0mm 14mm , clip=true]{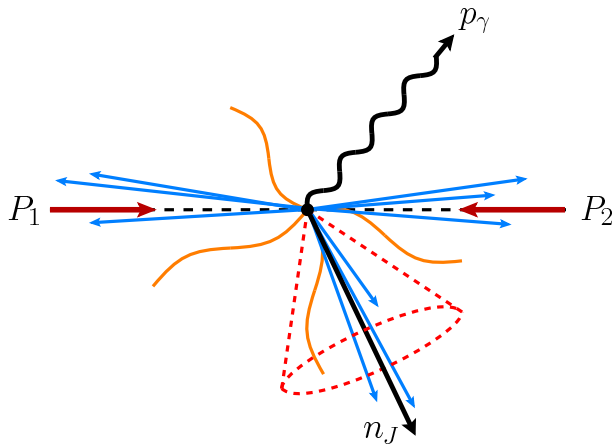}
    \caption{\label{fig:event3}Event topology of direct photon production.}
    \end{center}
\end{figure}

In this section we briefly review some kinematics of direct photon production
in a hadron collider and set up our notation.  We will follow the conventions
 in \cite{Becher:2009th}, and we summarize the most important
kinematic relationships.  Once the kinematic variables
are defined, we will define our jet definition and the specific observable
that we calculate in this paper.

\subsection{Hadronic and Partonic Kinematics}
\label{ssec:kinematics}

We will denote the momenta of the incoming hadrons as $P_1^\mu$ and $P_2^\mu$,
and the momentum of the photon as $p_\gamma^\mu$. The observable will be
categorized by the properties of the hard photon, namely its transverse
momentum $p_T$ and rapidity $y$. Formally, the results of this paper will be
most accurate when $p_T$ is near the {\it machine threshold} limit,
\be
p_T \sim p_T^{\rm max} = \frac{\ecm}{2 \cosh y}\, ,
\ee
where $\ecm = \sqrt{(P_1 + P_2)^2}$. A majority of events will have
photon transverse momenta that are far below the machine threshold limit.
However, as we will see, the calculation remains valid for transverse momenta
much less than machine threshold due a phenomenon called dynamical threshold
enhancement, discussed in Section \ref{sec:factthm}.

A schematic showing the event topology for the collision is given in
\fig{event3}. The hard process involves the collision between two initial state
partons, which have momentum $p_i^\mu = x_i P_i^\mu$ , where the $x_i$ is the
momentum fraction of the constituent parton and the subscript denotes the
hadron from which it came. There are two channels that are relevant at leading
order in $\alpha_s$, the annihilation channel, $q\bar{q} \to \gamma g$ and the Compton channel, $qg \to \gamma q$. It
is convenient to use the variables $w$ and $v$, given by
\be
x_1=\frac{1}{w}\frac{p_T}{\ecm v}e^y\;,
\hspace{1cm}
x_2=\frac{p_T}{\ecm (1-v)}e^{-y}\;.
\ee
The variable $v$ is related to the scattering angle between the photon and the
beam, $\theta^{\ast}$, in the partonic center of mass frame through
\be
v = \frac{1}{2}( 1  + \cos \theta^{\ast} )\, .
\ee
The other variable $w$ characterizes how much the (partonic) event deviates
from exact $2 \to 2$ kinematics due to additional QCD radiation in the initial
or final state. When $w = 1$, there is no additional radiation and the partonic
final state consists of a photon and either a quark in the case of $qg \to q
\gamma$ or a gluon in the case of $q \bar{q} \to g \gamma$.

When discussing the threshold limits, we will discuss the (total) hadronic
invariant mass
\be
    M_X^2=(P_1+P_2-p_\gamma)^2\;,
\ee
and the (total) partonic invariant mass
\be
    m_X^2=(x_1P_1+x_2P_2-p_\gamma)^2\;.
\ee
These give the invariant mass of all particles in the final state with the
photon's momentum removed from the hadronic or partonic final state,
respectively. The hadronic variable $M_X$ includes the beam remnant, whereas
the partonic version does not have any information about the beam, and
therefore $0\leq m_X \leq M_X$. In terms of the momentum fractions of the
partons and the photon transverse momentum and rapidity, the invariant masses
are
\be
  M_X^2=\ecm^2-2p_T \ecm\cosh y=\ecm^2\left( 1-\frac{p_T}{p_T^{\rm max}} \right)
\ee
and
\be
    m_X^2=\frac{p_T^2}{1-v}\frac{1-w}{w}\;.
\ee
In the partonic threshold limit, $w \to 1$, there is no phase space volume
for the initial or final state radiation, and so the the final state consists
of a single scattered parton and a photon. After removing the photon, only
a massless parton remains and so the partonic mass vanishes in this limit
($m_X\to0$). The partonic cross section is singular in this limit, and the
resulting singularities take the form of $\ln^2m_X \sim \ln^2(1-w)$. In the next section, we
will introduce a factorization theorem for the partonic cross section, that
will resum the logarithms of $m_X$ that appear as $\alpha_s^n\ln^m(1-w)$ in the
differential cross section.

\subsection{The observable}
\label{ssec:observable}

The observable we consider is the invariant mass of the hardest jet in events
with a high $p_T$ photon. Rather than
use one of the standard jet algorithms to define the jet mass, we use a simple
definition in order to make the analytical calculations more tractable. The jet
mass used in this paper is defined through the following procedure.
\begin{enumerate}
  \item Remove the photon from the event record.
  \item Cluster the event using the anti-$k_{T}$ jet algorithm.
  \item Select the jet with the hardest $p_T$, and find its 3-momentum
        $\mathbf{p}_J$. This defines the jet axis as $\hat{n}_J =
        \frac{\mathbf{p}_J}{ | \mathbf{p}_J |}$.
  \item Define $m_R$ to be the invariant mass of all of the radiation that lies
        within a cone of half-angle $R$ centered on $\hat{n}_J$.
\end{enumerate}
The observable we consider is $\frac{\rd \sigma}{\rd m_R}$, within some range of
$p_T$ and rapidity for the jet and photon. Our calculation is not sensitive
to the difference in $p_T$ between the jet and the photon, so we put a cut
only on the photon $p_T$, integrating inclusively over the $p_T$ of the jet. For
concreteness our numerical results will have central photons ($y=0$) with a
given $p_T$, integrated over jet rapidities between $-1$ and $1$.

Although our $R$ differs from a usual collider $R$, which is defined in terms
of rapidities and azimuthal angles, the difference is small for central jets
that dominate at large $p_T$. Our method of calculation, and our factorization
theorem, would also work for more standard jet algorithms, such as the $k_T$ algorithm.
However, most standard algorithms suffer from clustering effects \cite{Banfi:2005gj,
Delenda:2006nf, KhelifaKerfa:2011zu, Kelley:2012kj, Kelley:2012zs}
which are not completely understood. 
Extending the results of this paper to more standard jet
algorithms will be explored in future work.

To calculate the jet mass accurately in QCD, one would like the relevant events
to come from the hard process we are considering, namely a single photon
plus a single parton. However, generically, there can be contributions from
events with a hard photon and 2 or 3 jets. The simplest way to
ensure single jet kinematics is to use a $p_T$ veto on the second hardest jet.
This is easy to do experimentally or when using an event generator, but in QCD
introducing a new scale makes the theoretical calculation significantly more
complicated. In particular, non-global logarithms of the jet mass relative to
the veto scale will be introduced. Although these non-global logarithms might
be a small effect numerically, an advantage of the threshold resummation is
that one does not have to introduce an extra scale. Instead, the scale is
implicit in the out-of-jet soft radiation which is integrated over.

By expanding around the threshold limit and demanding large $p_T$ for the
photon, we can ensure that we are in the photon-plus-single-jet configuration
without introducing a new veto scale. When the photon's $p_T$ is on the
order of the hard scattering scale, the radiation outside of the hardest jet
must be soft, which prevents the formation of a second hard jet. By suitably
restricting the photon $p_T$, the events will contain a single recoiling jet
and we are safely in the regime where the factorization theorem is valid. To
put it another way, the fact that dynamical threshold enhancement has been
shown to work for the direct photon $p_T$ distribution in \cite{Becher:2009th}
indicates that the single jet configuration dominates. To be clear, the
threshold expansion does not obviate the need to deal with non-global
logarithms, as you would with a jet veto, but it does avoid having to deal with
an explicit scale.

\begin{figure}
    \begin{center}
    \includegraphics[scale=0.6, trim = 0mm 14mm 0mm 14mm , clip=true]{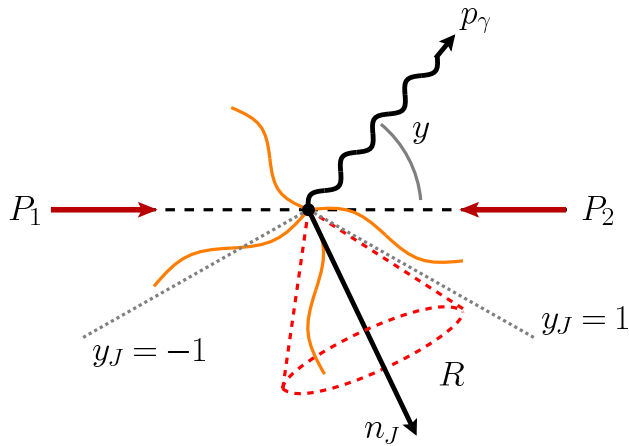}
    \caption{
    \label{fig:event4}
      The photon's momentum and rapidity is denoted by $p_\gamma$ and $y$, respectively.
      The jet cone of half angle $R$, centered around the direction of the jet $n_J = (1, \hat{n}_J)$.
      The restrictions on the jet rapidity $y_J$ is shown by the dashed gray line.  Soft
      radiation inside the jet contributes to $k_\in = n_J \cdot P_{X_s^{\in}}$, whereas
      the radiation outside contributes to $k_\out = n_J \cdot P_{X_s^{\out}}$
      }
    \end{center}
\end{figure}

At the partonic threshold limit, where the process only has one parton and a
photon in the final state, the $p_T$ and $y$ of the photon constrains the scattering angle
of the outgoing jet. To see this, consider momentum conservation along the
direction of the beam, taken to be the $z$ direction.
\bea
p_{J z} = (x_1 P_1 + x_2 P_2 - p_\gamma)_z
\eea
After using the definition of rapidity, $y = \frac{1}{2} \ln
\frac{E+p_z}{E-p_z}$, we can relate the rapidity of the jet $y_J$ to the $p_T$
and $y$ of the photon and to $x_{1,2}$ via
\bea
\sinh y_J = \frac{\ecm}{2p_T}(x_1-x_2)-\sinh y \;,
\eea
Since $0 < x_{1,2} < 1$, $y_J$ must lie within a range given by
\bea
-\frac{\ecm}{2p_T}-\sinh y
<
\sinh y_J
< \frac{\ecm}{2p_T}-\sinh y
\eea
In order to perform threshold resummation, we must be careful to keep $y_J$
small enough to ensure the jet does not enclose the beams.
This restriction is $\sinh y_J < \cot R$, and so if we take $|y_J|<1$, this implies that $R\lsim0.7$.

\section{Differential Cross Sections and Factorization Theorem}
\label{sec:factthm}

\begin{figure}
  \label{fig:pdfs}
    \begin{center}
    \includegraphics[scale=0.6, trim = 0mm 0mm 0mm 5mm , clip=true]{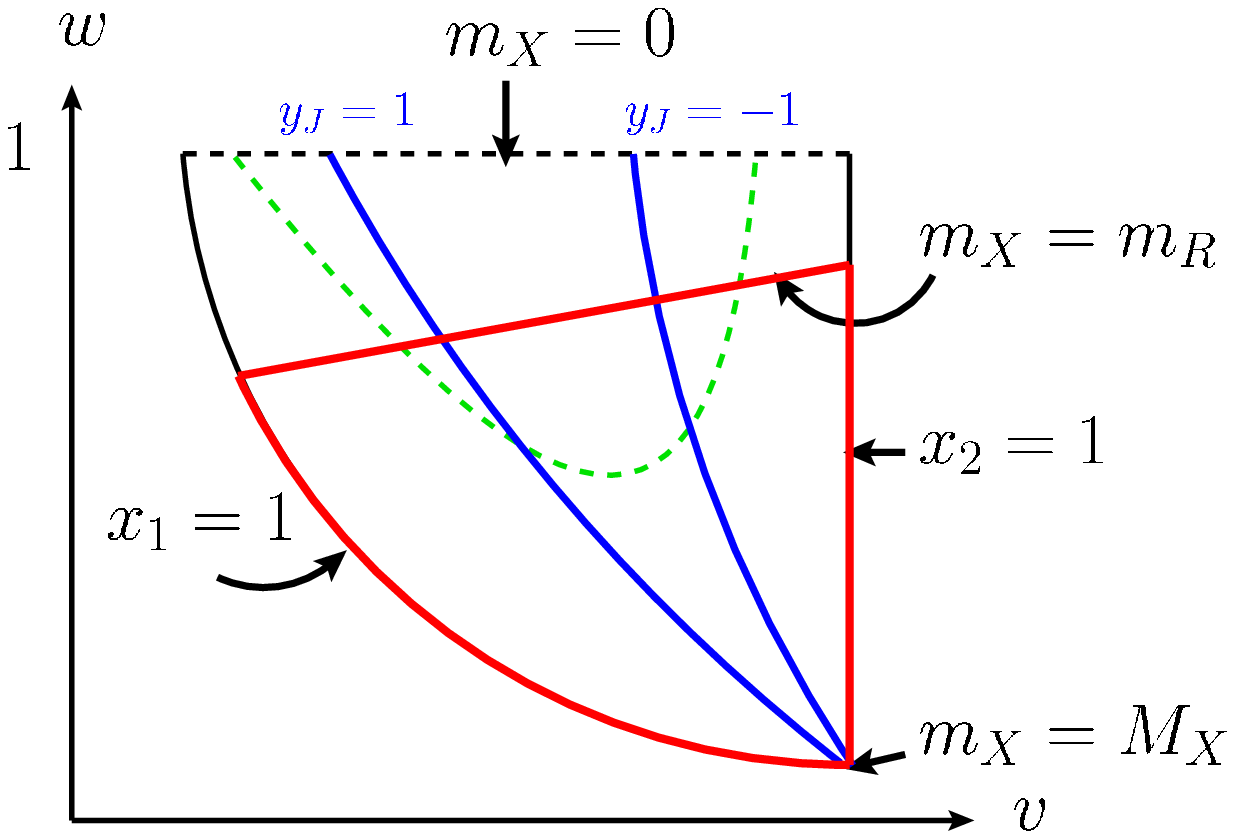}
    \includegraphics[scale=0.5, trim = 0mm 0mm 0mm 0mm , clip=true]{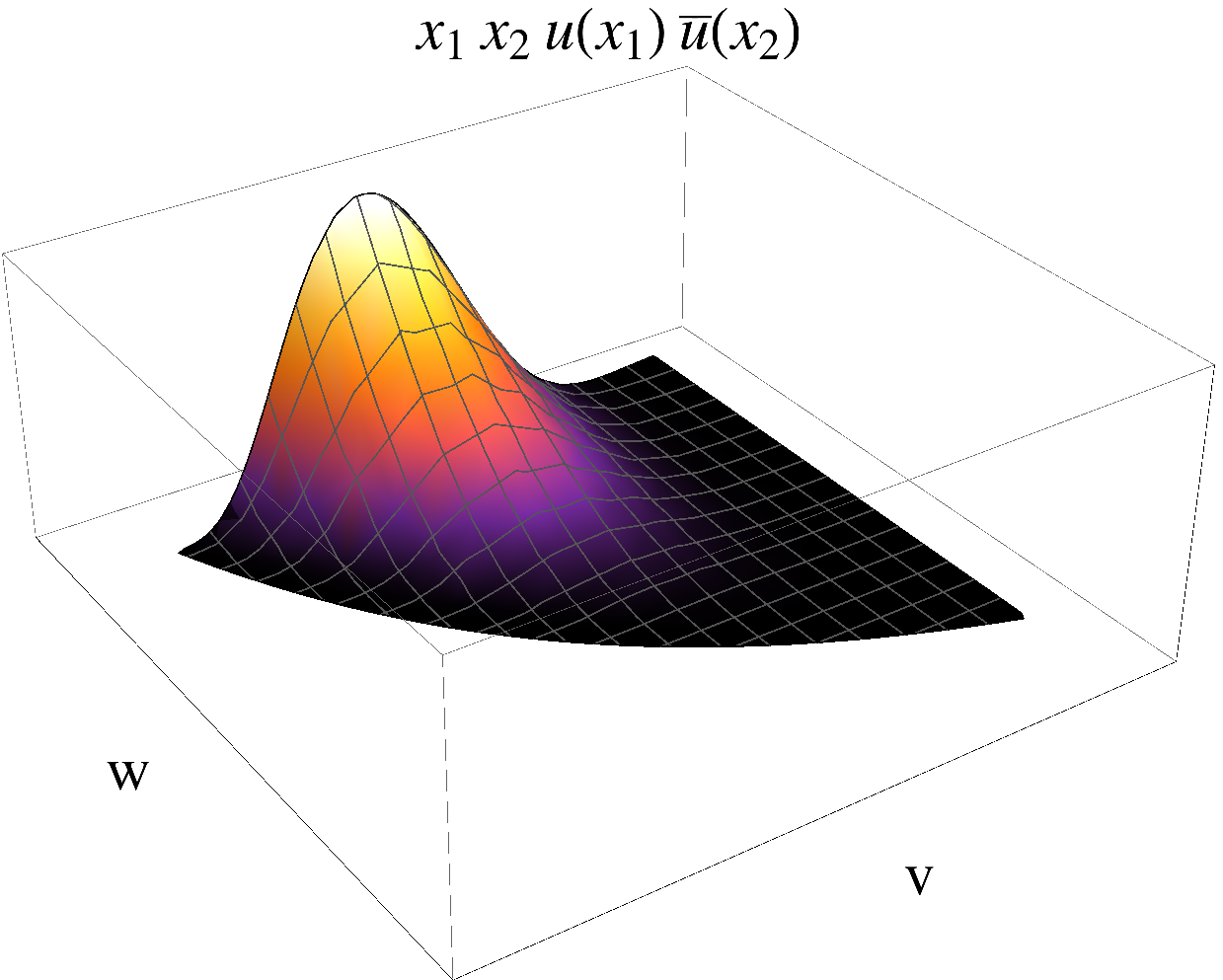}
    \caption{\label{fig:wv}
    The integration region $\mathcal{R}$ is shown by the red lines. When the jet
    rapidity is restricted, the region is further constrained to be inside the
    blue lines, shown here for $y_{\rm cut}=1$. The dashed green line outlines the
    region that most strongly contributes. This is demonstrated by the right
    hand plot, which shows the rapid fall off of $x_1 f_{u/P}(x_1) x_2 f_{\bar{u}/P}(x_2)$.
    }
  \end{center}
\end{figure}

The differential cross section for the process $N_1 + N_2 \to \gamma + X$ can be written in the form
\be
    \frac{ \rd^2\sigma}{ \rd p_T  \rd y \rd m_R^2}
    =\frac{2}{p_T}\sum_{ab}
    \int \!\!\!\! \int_{\mathcal{R}}\rd v \rd w~
    [x_1f_{a/N_1}(x_1,\mu)]~[x_2f_{b/N_2}(x_2,\mu)]
    ~\frac{ \rd^2\hat\sigma}{ \rd w  \rd v  \rd m_R^2}\;,
\ee
where $N_i$ are the incoming partons,  $f_{a/N_i}(x_i,\mu)$ is the probability distribution function for
finding parton $a$ in hadron $N_i$ with momentum fraction $x_i$, and
the sum is over the different partonic channels.
The region of integration $\mathcal{R}$ in the $w-v$ plane is given by
\bea
    \frac{1+m_R^2/p_T^2}{\frac{\ecm}{p_T}e^{-y}+m_R^2/p_T^2}
    < &v& < 1-\frac{p_T}{\ecm}e^{-y} \nn
    \frac{e^y}{v }
    \frac{p_T }{\ecm}
    < &w& <
    \frac{p_T^2}{ p_T^2+(1-v)m_R^2 } \nn
    \frac{e^y}{v(2\sinh y+2\sinh y_{\rm cut}+\frac{1}{1-v}e^{-y})}
    < &w& <
    \frac{e^y}{v(2\sinh y-2\sinh y_{\rm cut}+\frac{1}{1-v}e^{-y})}
    \,.
\eea
The first two constraints account for
$0 \leq x_i\leq1$ and $m_R^2\leq m_X^2$ , which is the region enclosed by the red lines in \fig{wv};
the third constraint comes from a finite rapidity interval for the jet: $-y_{\rm cut}\leq y_J \leq y_{\rm cut}$,
drawn as the blue lines in \fig{wv}.
The expression $\frac{ \rd^2\hat\sigma}{ \rd w  \rd v \rd m_R^2}$
gives the partonic cross section for $a + b \to \gamma + X$.  For a fixed $w$ and $v$,
the partonic cross section, at NLO, will diverge logarithmically as $m_R \to 0$,
and in this regime these logarithms can invalidate an expansion in $\alpha_s$.

The reason that the threshold expansion works away from threshold is
that the parton
distribution functions fall rapidly as the momentum fractions approach unity ($x_i \to 1$).
In \eqn{factorization}, the partonic differential cross section is weighted by
the product of the PDFs.  As a representative example, the right side of \fig{pdfs} shows a plot of the product
\be
x_1 f_{u/P}(x_1, \mu_f) x_2 f_{\bar{u}/P}(x_2, \mu_f)\;,
\ee
which would be necessary for process $u \bar{u} \to \gamma g$. In the plot, we have chosen $\ecm = 8$ TeV, $p_T = 2$
TeV, $y= 0$, and $\mu_f = \sqrt{\hat{s}} = \sqrt{x_1 x_2} \ecm$. The plot
clearly shows the extremely rapid fall off of the PDFs, and this demonstrates that,
for a fixed value of $m_R$, the region in $w-v$ space is weighted far more
heavily along the line given by $m_X = m_R$.

The hadronic invariant mass can be written as
\be
M_X^2 = \frac{m_X^2}{x_2} + \ecm^2[ (1-x_1)v + (1-x_2) (1-v)]  \;.
\ee
When $M_X \to 0$, this forces both $m_X \to 0$ and $x_{1,2} \to 1$; however,
the fall off the PDFs favor small $m_X$ even when $x_{i}$ is
not close to 1. Since $m_R < m_X$, this also means that small $m_R$ is favored.
The partonic cross section is singular in $m_R$, which
manifests itself as powers of $\ln m_R$, and for small
$m_R$, these logarithms are large and must be resummed. This enhancement of the
singular region for the partonic cross section means that resummation of $m_R$
is important, even when we are not near the machine threshold. This effect is
known as dynamic threshold enhancement~\cite{Appell:1988ie,Catani:1998tm,Becher:2007ty}.

\subsection{Factorization of the partonic cross section}
\label{ssec:fact}

When the jet invariant mass is small compared to the $p_T$, the jet is highly
collimated and the process can be described within the framework of the
Soft-Collinear Effective Theory (SCET)~\cite{Bauer:2000ew,Bauer:2000yr,Bauer:2001ct,Bauer:2001yt}. Rather than
give a formal derivation of the factorization formula as in \cite{Becher:2009th},
we will focus on how the factorization formula must be modified to become differential in the jet mass.

The partonic differential cross section can be calculated via
\be
\label{eq:factorization}
    \frac{ \rd^2\hat\sigma}{ \rd w  \rd v  \rd m_R^2}
    =w~{\tilde \sigma}(v)~H(p_T,v,\mu)\int_0^{\frac{m_R^2}{2E_J}}
    \frac{ \rd k}{2E_J}~J(m_R^2-(2E_J) k,\mu)~S(k,\frac{m_X^2-m_R^2}{2E_J},\mu)\;.
\ee
The three functions $H$, $J$ and $S$ are the hard, jet and soft functions,
which account for the physics associated with the different scales involved.
The quantity ${\tilde \sigma}(v)$ is defined through the leading
order QCD result via,
\be
    \frac{ \rd^2\hat\sigma}{ \rd w  \rd v  \rd m_R^2}
    ={\tilde \sigma_{ab}}(v)~\delta(m_R^2)~\delta(m_X^2-m_R^2)\;,
\ee
where the tree level parton cross sections are given by
\bea
\tilde{ \sigma}_{q\bar{q}}(v) =
  \pi \alpha_{\rm em} e_q^2\alpha_s(\mu)
  \frac{2C_F}{N_c} \left( \frac{v^2 + (1-v)^2}{1-v} \right) \nn
\tilde{ \sigma}_{qg}(v) =
  \pi \alpha_{\rm em} e_q^2\alpha_s(\mu)
  \frac{1}{N_c} ( 1 + (1-v)^2)\frac{v}{1-v} \;.
\eea
The hard function is given by the square of the Wilson coefficients that arise
when matching the QCD amplitude onto the effective field theory operators.
It describes the short distance physics at an energy scale $\mu_h \sim p_T$.
The jet function comes from integrating energetic modes that are collinear to
the jet direction that have virtuality comparable to $m_R$. In the threshold
limit $m_R \ll p_T R$, so the collinear radiation is insensitive to the
jet boundary. Thus we can use the same inclusive jet function as that in
\cite{Becher:2009th}. Finally, once the collinear modes are integrated out, we
are left with the soft modes, which are described by soft Wilson lines. The
soft function relevant for the jet mass calculation is different from the one
used in \cite{Becher:2009th}. As we will see, it depends separately on the
radiation inside and outside of the jet, and also on the jet algorithm.

The form of the factorization theorem can be understood from power counting in
the threshold region. In the partonic threshold limit, corresponding to the
limit $m_X \to 0$, radiation is constrained either to be collinear to recoiling
parton, which we call $p_J^\mu$, or to be soft, which we write as $k^\mu$. In
terms of these momentum regions, the partonic invariant mass takes the form
\be
\label{eq:mx}
    m_X^2=(p_J+k)^2= m_c^2+2k\cdot p_J + \cdots
\ee
up to terms of order $k^2\ll m_c^2$ where $m_c^2=p_J^2$ is the collinear mass,
whose distribution is given by the jet function.

The jet invariant mass $m_R$
only contains the part of radiation inside the cone. Since all the collinear
radiation is in the cone, we have only to split the soft radiation, which we
write as $k^{\mu}=k_\in^{\mu}+k_\out^{\mu}$, with $k_\in^{\mu}$ in the cone and
$k_\out^{\mu}$ out of the cone. Then,
\be
\label{eq:mR}
    m_R^2=(p_J+k_\in)^2= m_c^2+2k_\in\cdot p_J\;.
\ee
and so
\be
\label{eq:mx2}
    m_X^2 = m^2_R+2k_\out\cdot p_J\;.
\ee
Now, the jet momentum $p_J^\mu$ accounts for radiation that is collinear to recoiling
parton, and so it can be written as
\be
\label{eq:pJ}
p_J^{\mu} = E_J n^{\mu}_J + {\rm residual} \;.
\ee
where $n_J^\mu=(1,\hat{n}_J)$ is a lightlike vector that is directed along the
jet direction. The residual momentum gives a power suppressed contribution to
the jet mass compared to the leading component, and thus can be dropped. The
expressions for $m_X$ and $m_R$ then become
\bea
\label{eq:mXR}
    m_X^2 &=& m^2_R+ 2 E_J (n_J \cdot  k_\out) \nn
    m_R^2 &=& m^2_c+ 2 E_J (n_J \cdot k_\in) \;.
\eea
Therefore, the only component that enters the soft function is $k\equiv
n_J\cdot k$. These expressions explain the form of the factorization theorem,
Eq.(\ref{eq:factorization}), which can be written as
\bea
    \frac{ \rd^2\hat\sigma}{ \rd w  \rd v   \rd m_R^2}
    &=&w~{\tilde \sigma}(v)~H(p_T,v,\mu)
    \int  \rd k_\in \rd k_\out \rd m^2 J(m^2)S(k_\in,k_\out)
    \nn
    &&\qquad \times \delta(m_R^2-m^2-2E_Jk_\in)\delta(m_X^2-m_R^2-2E_Jk_\out)\;.
\label{eq:fact2}
\eea
Furthermore, we see why we need a soft function that depends on the projection
of the soft momentum on the jet direction separated into in-cone and out-of
cone components, which we denote $k_\in$ and $k_\out$.

The precise definition of the soft function we need is, for annihilation channel,
\be
\label{eq:soft_def1}
    S_{q\bar q}(k_\in,k_\out)
    =\frac{1}{C_F N_c}
    \sum_{X_s}
    | \langle X_s|{\bf T}\Big[Y_1^\dagger Y_Jt^aY_J^\dagger Y_2(0)\Big]|0\rangle|^2
    \delta(n_J\cdot P_{X_s^\in}-k_\in)
    \delta(n_J\cdot P_{X_s^\out}-k_\out)\;,
\ee
and for Compton channel
\be
\label{eq:soft_def2}
    S_{qg}(k_\in,k_\out)
    =\frac{1}{C_F N_c}
    \sum_{X_s}
    |\langle X_s|{\bf T}\Big[Y_1^\dagger Y_2t^aY_2^\dagger Y_J(0)\Big]|0\rangle|^2
    \delta(n_J\cdot P_{X_s^\in}-k_\in)
    \delta(n_J\cdot P_{X_s^\out}-k_\out)\;,
\ee
where $P_{X_s^\in}$ and $P_{X_s^\out}$ denote the total momentum of the soft
radiation propagating inside and outside the jet, respectively. Also,
$Y_i$ is a soft Wilson line directed along the $n_i = (1, \hat{n}_i)$ direction:
\be
Y^{\dagger}_{i}(x) = P \exp \left( ig \int_0^{\infty} ds\ n_i \cdot A_s(n_i s + x) \right) \;,
\ee
where $i = 1,2,J$ and $P$ denotes path ordering and $A_s = A_s^a T^a$
are soft gauge fields in the fundamental representation.

\subsection{One-loop soft function}
\begin{figure}
    \begin{center}
    \includegraphics[scale=0.23, trim = 2mm 0mm 0mm 0mm , clip=true]{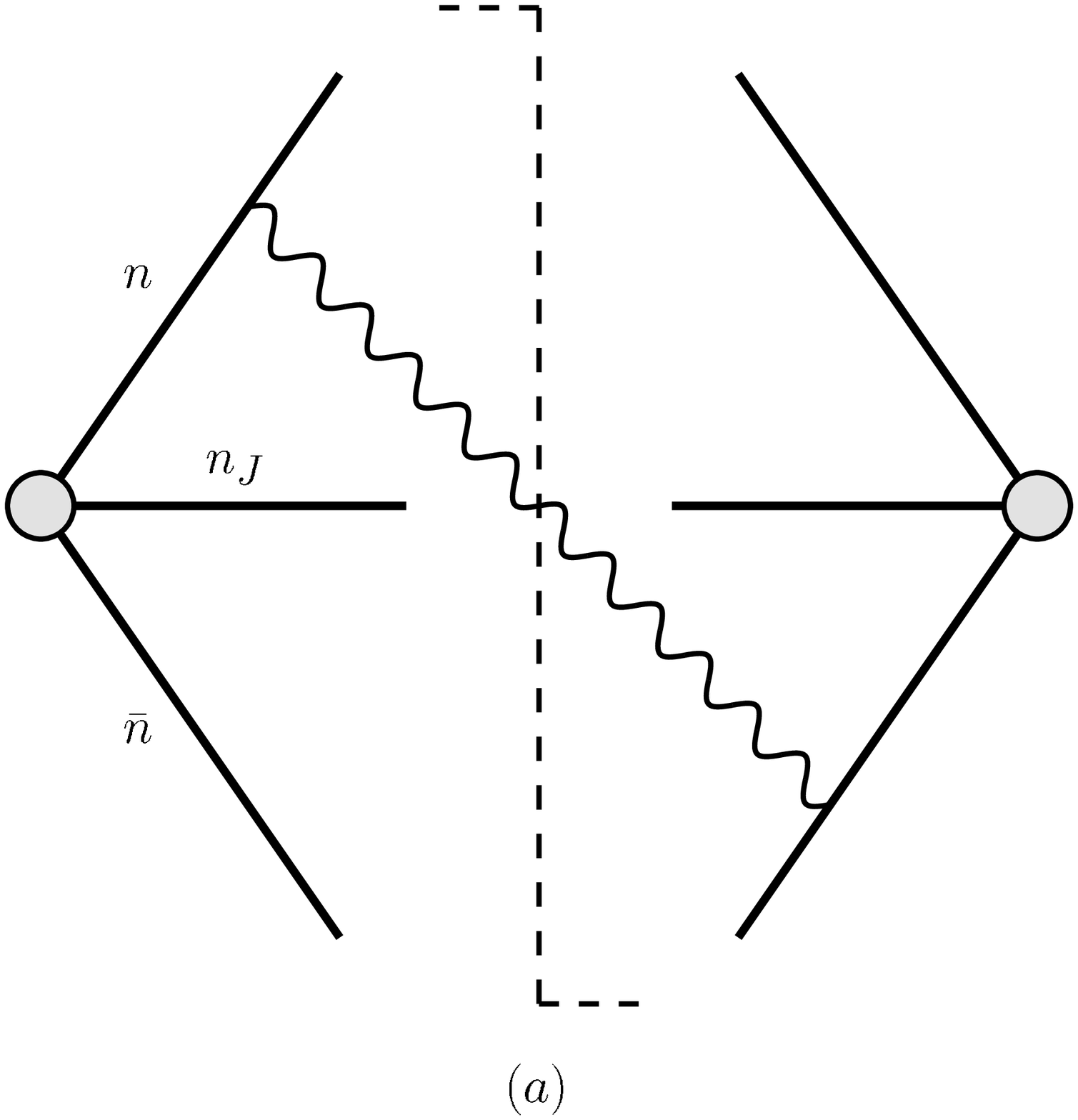}
    \hspace{3mm}
    \includegraphics[scale=0.23, trim = 2mm 0mm 0mm 0mm , clip=true]{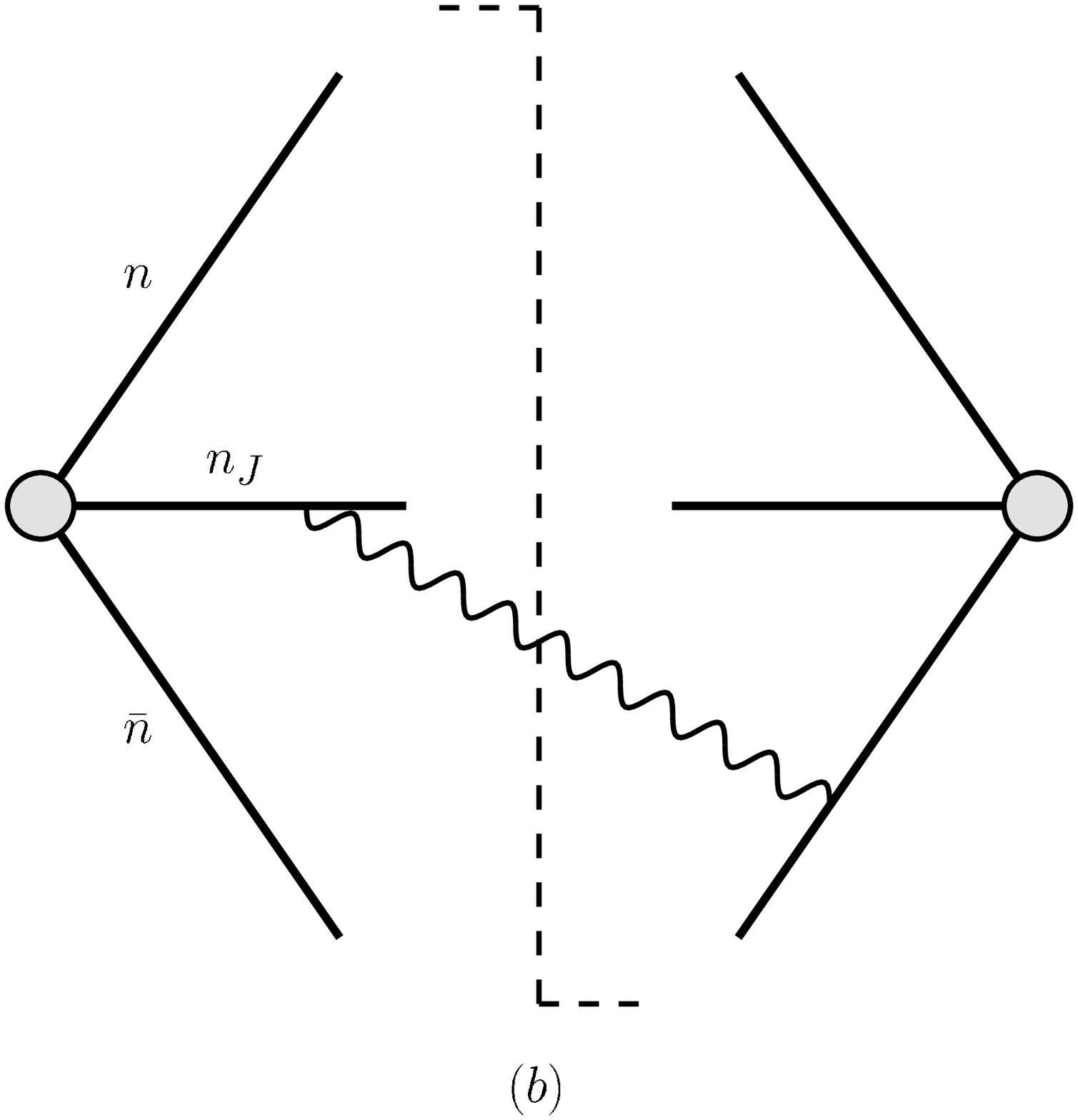}
    \hspace{3mm}
    \includegraphics[scale=0.23, trim = 2mm 0mm 0mm 0mm , clip=true]{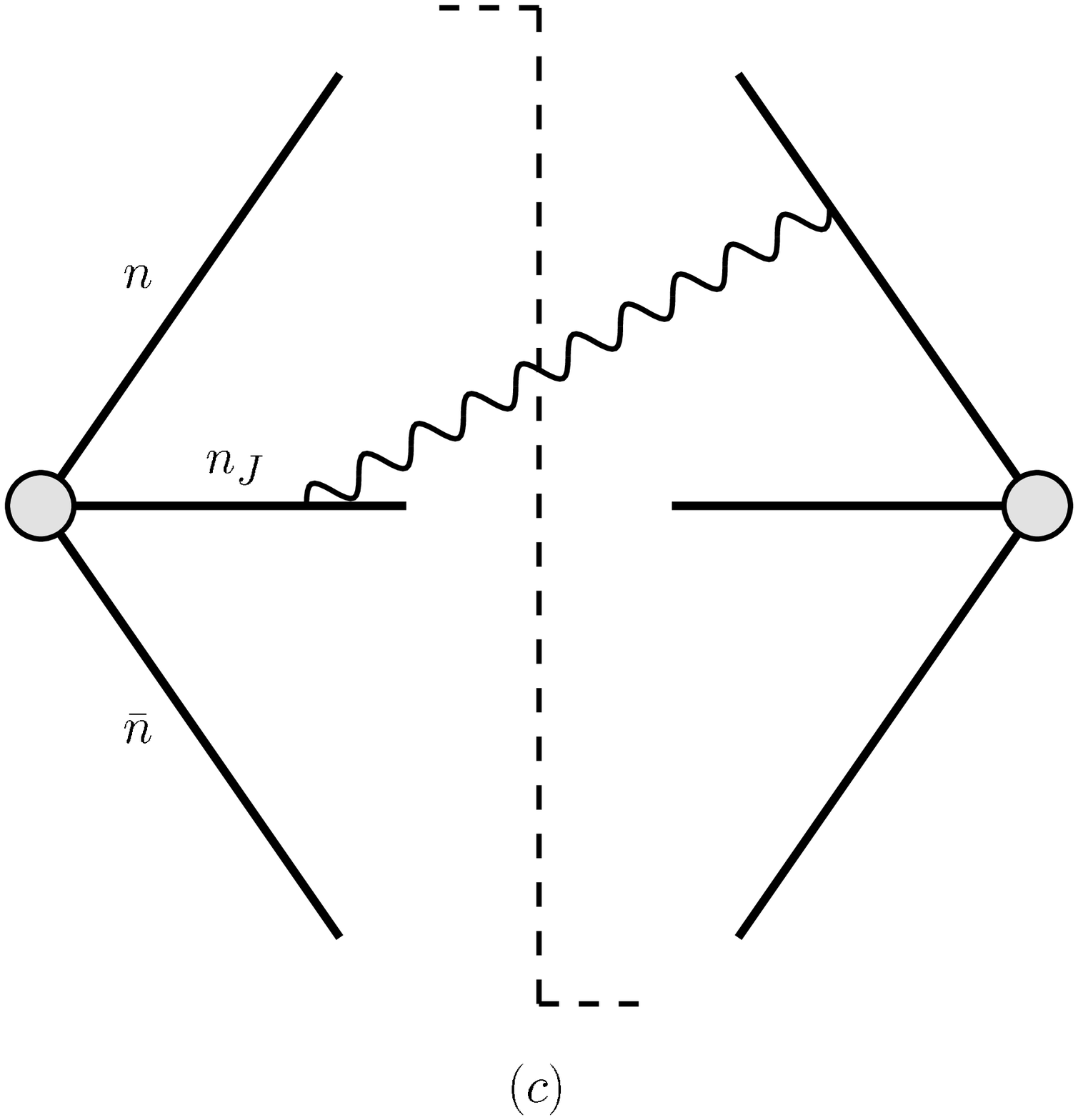}
    \caption{
    \label{fig:cuts}
      The non-vanishing diagrams that contribute to the one-loop soft function.
    }
    \end{center}
\end{figure}

We now calculate the required soft function to order $\alpha_s$.
The diagrams needed for the calculation are represented in \fig{cuts}.
The virtual diagrams are not shown as they give a scaleless contribution when
using dimensional regularization to regulate both the ultraviolet and infra-red divergences.
The non-vanishing diagrams for the $q\bar q\rightarrow g\gamma$ channel give
\bea
\label{eq:soft_1loop_qq}
&&S_{q\bar q}(k_\in,k_\out, \mu)
=
  \frac{8\pi \alpha_s}{(2\pi)^{d-1}}
  \Big(\frac{\mu^2e^{\gamma_E}}{4\pi}\Big)^{\epsilon}
  \int d^dq \
  \delta(q^2)\theta(q_0)
  \mathcal{M}( k_\in , k_\out; q)
\nn
&&\times
  \Big[
    \Big(C_F-\frac{1}{2}C_A\Big)\frac{n_1\cdot n_2}{(n_1\cdot q)(n_2\cdot q)}
    +\frac{1}{2}C_A\frac{n_J\cdot n_1}{(n_J\cdot q)(n_1\cdot q)}
    +\frac{1}{2}C_A\frac{n_J\cdot n_2}{(n_J\cdot q)(n_2\cdot q)}
    \Big]\;,
\eea
where $q^+=n_J\cdot q$,  and $q^-={\bar n}_J\cdot q$ with $\nbar_J = (1, -\hat{n}_J)$,
and the measurement function is given by
\bea
\mathcal{M}(k_\in , k_\out; q)
&=&\Theta\Big(\frac{q^--q^+}{q^-+q^+}-\cos\thR\Big)\delta(k_\in-q^+)\delta(k_\out) \nn
&&  +\Theta\Big(\cos\thR-\frac{q^--q^+}{q^-+q^+}\Big)\delta(k_\in)\delta(k_\out-q^+) \;.
\eea
Similarly, for the $qg\rightarrow q\gamma$ channel we have
\bea
\label{eq:soft_1loop_qg}
&&S_{qg}(k_\in,k_\out, \mu)
=
  \frac{8\pi \alpha_s}{(2\pi)^{d-1}}
  \Big(\frac{\mu^2e^{\gamma_E}}{4\pi}\Big)^{\epsilon}
  \int d^dq \
  \delta(q^2)\theta(q_0)
  \mathcal{M}( k_\in , k_\out; q)
\nn
&&\times
  \Big[
    \frac{1}{2}C_A\frac{n_1\cdot n_2}{(n_1\cdot q)(n_2\cdot q)}
    +\Big(C_F-\frac{1}{2}C_A\Big)\frac{n_J\cdot n_1}{(n_J\cdot q)(n_1\cdot q)}
    +\frac{1}{2}C_A\frac{n_J\cdot n_2}{(n_J\cdot q)(n_2\cdot q)}
  \Big].
\eea
The soft function will depend on the angle $\thR$ and the rapidity of the jet
$y_J$.  For convenience we define
\bea
r
= \tan \frac{\thR}{2} \hspace{1cm}
\beta
= \exp(-y_J)
\eea
With these definitions, the soft functions associated with the two channels
at order $\alpha_s$ are
\bea
\label{eq:soft_full}
    S_{q\bar q}(k_\in,k_\out,\mu)
    &=&
    \delta(k_\in)\delta(k_\out)
    +2\Big(\frac{\alpha_s}{4\pi}\Big)
    \Big\{
      \delta(k_\in)\delta(k_\out)
      \Big(C_F-\frac{1}{2}C_A\Big)
      \Big[\ln^2\frac{(1+\beta^2)^2}{\beta^2}-\frac{\pi^2}{6}\Big]\nn
      &+&
      2\frac{\delta(k_\out)}{k_\in}
      \Big[  -2C_A\ln\Big(\frac{k_\in}{\mu}\frac{1+\beta^2}{\beta}\Big)
             +c_{q\bar q}
      \Big]
      \nn
      &+&
      2\frac{\delta(k_\in)}{k_\out}
      \Big[4C_F\ln\Big(\frac{k_\out}{\mu}\frac{1+\beta^2}{\beta}\Big)-c_{q\bar q}
      \Big]
    \Big\} \;,
\eea
and
\bea
    S_{qg}(k_\in,k_\out,\mu)
    &=&
    \delta(k_\in)\delta(k_\out)
    +2\Big(\frac{\alpha_s}{4\pi}\Big)
    \Big\{
      \delta(k_\in)\delta(k_\out)
      \Big(\frac{1}{2}C_A\Big)
      \Big[
        \ln^2\frac{(1+\beta^2)^2}{\beta^2}-\frac{\pi^2}{6}
      \Big]\nn
      &+&
      2\frac{\delta(k_\out)}{k_\in}
      \Big[
        -2C_F\ln\Big(\frac{k_\in}{\mu}\frac{1+\beta^2}{\beta}\Big)+c_{qg}
      \Big]
      \nn
      &+&
      2\frac{\delta(k_\in)}{k_\out}
      \Big[
        (2C_F+2C_A)\ln\Big(\frac{k_\out}{\mu}\frac{1+\beta^2}{\beta}\Big)-c_{qg}
        \Big]\Big\}\;,
\eea
where
\bea
\label{eq:constants}
    c_{q\bar q}
    &=&-C_F\ln \left[ \frac{ (\beta^2-r^2)(1 - r^2\beta^2)}{\beta^2 r^4}  \right]
    -C_A\ln\frac{\beta^2}{(1+\beta^2)^2}+(2C_F-C_A)\ln \frac{1+r^2}{r^2}\nn
    c_{qg}
    &=&\Big[-C_F\ln \frac{\beta^2 - r^2}{r^2 \beta^2} -C_A\ln \frac{1-r^2 \beta^2}{r^2 \beta^2} \Big]
    -C_F\ln\frac{\beta^2}{(1+\beta^2)^2}+C_A\ln\frac{1+r^2}{r^2} \;.
\eea
The logarithmic divergences involving $r$ and $\beta$ arise when the jet cone encloses
the collinear radiations from one of the incoming beams.

Since the factorization theorem contains convolutions between the jet
and soft functions, it is easier to perform the calculation in Laplace space.
The Laplace transformed soft function is defined as
\bea
    {\tilde s}(\kappa_\in,\kappa_\out,\mu)
    &=&
    \int_0^\infty dk_\in\int_0^\infty dk_\out
    \exp\Big(-\frac{k_\in}{\kappa_\in e^{\gamma_E}}\Big)\exp\Big(-\frac{k_\out}{\kappa_\out e^{\gamma_E}}\Big)
    S(k_\in, k_\out,\mu) \;. \nn
\eea
Our result is then
\bea
    {\tilde s}_{q\bar q}(L_1,L_2,\mu)
    &=&1+\Big(\frac{\alpha_s}{4\pi}\Big)
      \Big[
        4(2C_FL_2^2-C_AL_1^2)+4(L_1-L_2)c_{q\bar q}+\Big(C_F-\frac{1}{2}C_A\Big)\pi^2
      \Big] \nn
    {\tilde s}_{qg}(L_1,L_2,\mu)
    &=&1+\Big(\frac{\alpha_s}{4\pi}\Big)
      \Big[
        4((C_F+C_A)L_2^2-C_FL_1^2)+4(L_1-L_2)c_{qg}+\Big(\frac{1}{2}C_A\Big)\pi^2
      \Big]\;, \nn
\eea
where  $L_1=\ln\Big(\frac{1+\beta^2}{\beta}\frac{\kappa_\in}{\mu}\Big)$
and $L_2=\ln\Big(\frac{1+\beta^2}{\beta}\frac{\kappa_\out}{\mu}\Big)$.

The renormalization group equations satisfied by the soft function in the two channels are
\bea
\label{eq:gamma_soft}
\frac{d \tilde{s}_{q\bar q}}{d \ln\mu}
    &=& \Big[
      2C_A\gamma_{\rm cusp}
      \ln\Big(\frac{1+\beta^2}{\beta}\frac{\kappa_\in}{\mu}\Big)
        -4C_F\gamma_{\rm cusp}
      \ln\Big(\frac{1+\beta^2}{\beta}\frac{\kappa_\out}{\mu}\Big)
        -2\gamma^{S_{q{\bar q}}}
    \Big]
    \tilde{s}_{q\bar q }
\nn
    \frac{d\tilde s_{qg}}{d\ln\mu}
    &=&
     \Big[
      2C_F\gamma_{\rm cusp}
      \ln\Big(\frac{1+\beta^2}{\beta}\frac{\kappa_\in}{\mu}\Big)
      -2(C_F+C_A)\gamma_{\rm cusp}
      \ln\Big(\frac{1+\beta^2}{\beta}\frac{\kappa_\out}{\mu}\Big)
      -2\gamma^{S_{qg}}
     \Big]
     \tilde{s}_{qg} \;. \nn
\eea
where the expressions for $\gamma_{\rm cusp}$, $\gamma^{S_{q\bar{q}}}$,
$\gamma^{S_{qg}}$ are given to order $\alpha_s^2$ in \cite{Becher:2009th}
and are known to order $\alpha_s^3$.

To check that the factorization theorem in \eqn{factorization} is correct,
we have confirmed that the $\mu$-dependence of the hard, jet,
soft functions and PDFs near $x=1$ cancel when combined. We also compare the singular part of the full QCD
distribution at high $p_T$ which we compute using the numerical package {\sc mcfm} to the distribution produced by combining the hard, jet and soft functions at fixed order. This is shown in Figure \ref{fig:mcfm}.

Although the factorization theorem is correct, it does not automatically guarantee that we
can resum all of the logarithms
of $m_R$. Recall that the soft function in the factorization theorem in
\cite{Becher:2009th} was fully inclusive and had dependence only on the
projection of the soft momentum in the jet direction. For the inclusive
photon $p_T$ spectrum, such a one-scale soft function is natural, since the
observable only depends on a single scale, $p_T$. In contrast, the soft
functions given in \eqns{soft_def1}{soft_def2} depend on two physical scales
$k_\in, k_\out$ as well as on the renormalization scale $\mu$. With multiple
soft scales, there may not be a natural choice for $\mu$, and therefore the
renormalization group evolution may not resum all of the large logarithms.
Indeed, in a analogous jet mass calculation for an $e^+e^-$ collider, it was
demonstrated first numerically in \cite{Kelley:2011tj} and then analytically 
in~\cite{Kelley:2011aa} that the singularity structure of QCD was not completely
reproduced by expanding the resummed result for jet thrust. One expects the same difficulty
here.

The problem is that our soft function anomalous dimension does not depend on
jet size $R$. This is simply because the hard and jet functions as well as
the PDFs are the same as in the inclusive case and do not have a jet size
dependence. However, there are large logarithms in full QCD that could only
be predicted from an $R$-dependent anomalous dimension. To make this point more
concrete, consider the singular part of the cross section calculated to
${\mathcal{O}}(\alpha_s)$ using SCET in the annihilation channel:
\be
    \frac{d\hat\sigma_{q\bar q}}{ \rd w  \rd v  \rd m_R^2}
    =w{\tilde \sigma_{q\bar q}}(v)
    \delta(m_X^2-m_R^2)
    \frac{\alpha_s(\mu)}{4\pi}
    \frac{-2C_A\Gamma_0\ln\frac{m_R^2}{p_T\mu}+
   {c_{q\bar q}(r, \beta)}
      \Gamma_0+\Gamma^{J_g}_0\ln \frac{m_R^2}{\mu^2}+\gamma^{J_g}_0}{m_R^2}
\ee
Since SCET contains all of the singular contributions from full QCD, this
will agree with the full one-loop QCD result expanded at small $m_R$. However,
we should also be able to predict these $\frac{1}{m_R^2}$ terms only using
the anomalous dimensions, since they should be part of the resummation at
NLL level. Since the coefficient of $\frac{1}{m_R^2}$ depends on $r = \tan
R/2$ through the ${c_{q\bar q}(r,\beta)}$ term, this is
impossible for $R$-independent anomalous dimensions. Thus, without
further insight, resummation in SCET will undoubtedly miss these logarithmic
corrections.

\section{Refactorization of the Soft Function}
\label{sec:soft_fact}

The soft function depends on the two scales $k_\in$ and $k_\out$. By
explicit calculation we found it has logarithmic dependence on both of
the scales divided by $\mu$ already at one loop. Since there is not a
single scale associated with the soft function, we should not expect
that a single choice of $\mu$ will give the soft function a controlled
perturbative expansion. In addition, at two loops, one generically expects a
complicated dependence on $k_\in/k_\out$ in the soft function, as was found
in~\cite{Kelley:2011ng,Kelley:2011aa}, representative of the non-global
structure. The problem, from an effective theory point of view, is that there
are two modes with different characteristic scaling described by the same
object. Ideally, we could factorize the soft function into separate objects
which depend only on $k_\in /\mu$ and $k_\out/\mu$. However, it does not appear
that the soft function exactly factorizes in that way.

In the absence of a full refactorization, we will proceed in a conservative
fashion following methods of traditional resummation \cite{Kidonakis:1998nf,
Berger:2003iw, Sterman:1995fz, Collins:1989gx, kidonakis:1998bk,
kidonakis:1997gm}. To that end, we can at least isolate part of the soft
function whose scaling we already know, namely the part associated with
radiation that goes entirely into the jet, which will at least contain all of
the soft-collinear divergences. To do so, we will define an auxiliary soft
function that only depends
on $k_\in$, reminiscent of eikonal jet functions in \cite{Kidonakis:1998nf}.
There are many ways to do this. A simple choice is to define the auxiliary
soft function the same way we define the 2-scale soft function, but inclusive
over the out-of-jet radiation. That is, we take
\be
\label{eq:soft_aux_jet_qq}
    \bS_{q\bar q}(k_\in)
    =\frac{1}{C_F N_c}
    \sum_{X_s}
    | \langle X_s|{\bf T}\Big[Y_1^\dagger Y_Jt^aY_J^\dagger Y_2(0)\Big]|0\rangle|^2
    \delta(k_\in - n_J\cdot P_{X_s^\in}) \;,
\ee
for the annihilation channel, and similarly for the Compton channel.
We call this a {\bf regional soft function}.
The regional soft function was deliberately chosen to have the same color structure as
the original soft function.

At one-loop the regional soft function is given by \eqns{soft_1loop_qq}{soft_1loop_qg}
with the measurement function replaced by
\bea
\mathcal{M}(k_\in ; q)
&=&\Theta\Big(\frac{q^--q^+}{q^-+q^+}-\cos R\Big)\delta(k_\in-q^+) \;.
\eea
At order $\alpha_s$, we find
\bea
\label{eq:soft_aux}
    \bS_{q\bar q}(k_\in,\mu)
    &=&
    \delta(k_\in)
    +2\Big(\frac{\alpha_s}{4\pi}\Big)\Big\{\delta(k_\in)\Big(-\frac{1}{2}C_A\Big)
    \Big[
      \ln^2\frac{(1+\beta^2)^2}{\beta^2}-\frac{\pi^2}{6}\Big]\nn
      &+&2\frac{1}{k_\in}\Big[-2C_A\ln\Big(\frac{k_\in}{\mu}\frac{1+\beta^2}{\beta}\Big)
      +c_{q\bar q}
    \Big]
    \;,
    \nn
    \bS_{qg}(k_\in,\mu)
    &=&
    \delta(k_\in)
    +2\Big(\frac{\alpha_s}{4\pi}\Big)\Big\{\delta(k_\in)\Big(-\frac{1}{2}C_F\Big)
    \Big[
      \ln^2\frac{(1+\beta^2)^2}{\beta^2}-\frac{\pi^2}{6}\Big]\nn
      &+&2\frac{1}{k_\in}\Big[-2C_F\ln\Big(\frac{k_\in}{\mu}\frac{1+\beta^2}{\beta}\Big)
      +c_{qg}
    \Big]
    \;,
\eea
where $c_{q\bar{q}}$ and $c_{qg}$ are given in \eqn{constants}.
The anomalous dimensions at order $\alpha_s$, in Laplace space,  are
\bea
\label{eq:gamma_aux}
\frac{ d}{ d\ln\mu}
    \tilde{s}_{q\bar{q}}(\kappa_\in ,\mu)
    &=& \frac{\alpha_s}{4\pi}
    \Big[
    2 C_A \Gamma_{0}
    \ln \left( \frac{1+\beta^2}{\beta} \frac{\kappa_\in}{\mu} \right)
    - 4 c_{q \bar{q}}(r,\beta)
    \Big]
    \tilde{s}_{q\bar{q}}(\kappa_\in ,\mu) \nn
\frac{ d}{ d\ln\mu}
    \tilde{s}_{qg}(\kappa_\in ,\mu)
    &=& \frac{\alpha_s}{4\pi}
    \Big[
    2 C_F \Gamma_{0}
    \ln \left( \frac{1+\beta^2}{\beta} \frac{\kappa_\in}{\mu} \right)
    - 4 c_{qg}(r,\beta)
    \Big]
    \tilde{s}_{qg}(\kappa_\in ,\mu)\;,
\eea

The dependence of the soft function on $k_\out$ is still encoded in the full soft function,
so we define the {\bf residual soft function} to be what remains after
dividing the full soft function by the regional soft function.
\be
\rS(k_\in, k_\out, \mu) = \frac{S(k_\in, k_\out, \mu)}{\bS(k_\in, \mu) }
\ee
At order $\alpha_s$, the contribution from radiation going inside the jet is identical
between the full soft function and the regional soft function. When the radiation
goes outside of the jet, it contributes to $k_\out$ in the full contribution, however
its contribution to the regional soft function vanishes since the integral is scaleless.
Thus the residual soft function at order $\alpha_s$ only has $k_\out$ dependence:
\bea
\label{eq:soft_residual}
    {\rS}_{q\bar q}(k_\in, k_\out,\mu)
    &=&
    \delta(k_\out)
    +2\Big(\frac{\alpha_s}{4\pi}\Big)\Big\{\delta(k_\out)\Big(C_F\Big)
    \Big[\ln^2\frac{(1+\beta^2)^2}{\beta^2}-\frac{\pi^2}{6}\Big]\nn
    &+&
    2\frac{1}{k_\out}\Big[4C_F\ln\Big(\frac{k_\out}{\mu}\frac{1+\beta^2}{\beta}\Big)-c_{q\bar q}\Big]\Big\}\nn
    {\rS}_{qg}(k_\in,k_\out,\mu)
    &=&
    \delta(k_\out)
    +2\Big(\frac{\alpha_s}{4\pi}\Big)\Big\{\delta(k_\out)
    \Big(\frac{1}{2}C_F+\frac{1}{2}C_A\Big)\Big[\ln^2\frac{(1+\beta^2)^2}{\beta^2}-\frac{\pi^2}{6}\Big]\nn
    &+&
    +2\frac{1}{k_\out}\Big[(2C_F+2C_A)\ln\Big(\frac{k_\out}{\mu}\frac{1+\beta^2}{\beta}\Big)-c_{qg}\Big]\Big\}\;.
\eea
Thus, the one-loop anomalous dimensions of the residual soft function depends only
on the Laplace conjugate variable for $k_\out$, $\kappa_\out$:
\bea
\label{eq:gamma_residual}
\frac{ d}{ d\ln\mu}
    \tilde{s}_{rq\bar{q}}(\kappa_\out, \mu)
    &=& \frac{\alpha_s}{4\pi}
    \Big[
    -4 C_F \Gamma_{0}
    \ln \left(\frac{1+\beta^2}{\beta} \frac{\kappa_\out}{\mu} \right)
    + 4 c_{q \bar{q}}(r,\beta)
    \Big]
    \tilde{s}_{rq\bar{q}}(\kappa_\out,\mu) \nn
\frac{ d}{ d\ln\mu}
    \tilde{s}_{rqg}(\kappa_\out ,\mu)
    &=& \frac{\alpha_s}{4\pi}
    \Big[
    -2 (C_F+C_A) \Gamma_{0}
    \ln \left(\frac{1+\beta^2}{\beta} \frac{\kappa_\out}{\mu} \right)
    + 4 c_{qg}(r,\beta)
    \Big]
    \tilde{s}_{rqg}(\kappa_\out ,\mu)\;, \nn
\eea

Combining these results, we see that we have actually factorized the soft
function at one-loop, since the regional and residual soft functions each
depend on only one scale. At two-loops and higher, the regional soft function
will still only depend on $k_\in$, however the residual soft function
will have both $k_\in$ and $k_\out$ dependence. Thus, at least there should
be a natural scale which minimizes the large logarithms in $\bS(k_\in,\mu)$.
In particular, this scale should correspond to the ``see-saw'' scale $\mu_s =
\mu_j^2/\mu_h$ for the soft radiation associated with a particular collinear
direction~\cite{Schwartz:2007ib}. Such a scale assignment is implicit in much
of the work on traditional resummation, and here we have made it explicit
using SCET.

For this refactorization to make any difference, we have to allow the residual
soft function scale to be different from the scale for the regional soft
function. So we write
\be
S(k_\in, k_\out, \mu) =
\hat{\Pi}(\mu,\mu_\in)\bS(k_\in, \mu_\in)
\Pi_r(\mu,\mu_\out)\rS(k_\in, k_\out, \mu_\out)
\label{eq:refact}
\ee
where the evolution kernels $\hat{\Pi}(\mu,\mu_\in)$ and $\Pi_r(\mu,\mu_\out)$ evolve
their respective soft functions from $\mu$ to separate scales. In \sec{results}, we
investigate how choosing two soft scales affects the jet mass distribution
by comparing the calculation with and without such a scale separation (see
\fig{500a4}).

Unfortunately, since the residual soft function will depend on both $k_\in$ and $k_\out$ beyond
one-loop, our refactorization is only approximate. At two-loops, the $k_\in/k_\out$ dependence of
the residual soft function is due to non-global structure.
When the scales are widely separated, $k_\in \ll k_\out$ or vice
verse, the non-global structure takes the form of a logarithm of the ratio
of these scales.  There have been many studies of non-global logarithms (NGLs)
\cite{Dasgupta:2001sh,Dasgupta:2002bw,Banfi:2002hw,Appleby:2002ke,
Banfi:2010pa,Kelley:2011ng,Hornig:2011iu,KhelifaKerfa:2011zu,Kelley:2011aa}.
We will not attempt to calculate the non-global structure for this observable,
which would require a two-loop calculation, instead
we will estimate their effects and discuss their phenomenological significance
in \sec{NGL}.

\subsection{Comparing with pQCD result}

Calculating the jet mass distribution at hadron collider has
been done by using traditional pQCD resummation technique
\cite{Li:2011hy,Li:2012bw,Dasgupta:2012hg}. It's interesting to compare
the pQCD results and the SCET results obtained in this paper. In
Refs.~\cite{Li:2011hy,Li:2012bw}, the jet mass distribution is described
by a process independent quark or gluon jet function, depending on the
partonic origin of the jet. These pQCD jet functions are calculated to
$\mathcal{O}(\alpha_s)$ explicitly, and resumed to all orders at NLL level
by solving an evolution equation in Mellin space. In their approach, both
the hard-collinear mode and soft mode contribute to the jet function, in
contrast to the SCET approach, where jet function is defined with zero-bin
subtraction \cite{Manohar:2006nz}. Furthermore, the jet function defined in
Refs.~\cite{Li:2011hy,Li:2012bw} is independent of underlying partonic process,
and the corresponding jet mass distribution misses the contribution from soft
large angle radiation between different colored partons. Therefore a simple
connection between their jet function and our SCET results can not be made.

To make progress comparing the different formalisms, it is helpful to consider
the $R\to 0$ limit. In this limit, a process independent combination of jet
and soft function can be defined in SCET. For example, for the annihilation
channel, we have
\begin{equation}
\label{eq:jetsmallR}
  J_g^{SCET}(m^2_R,\mu) = \lim_{R\to 0}\int dp^2 dk_{\rm in} \, J_g(p^2,\mu) S_{q\bar{q}}(k_{\rm in},\mu)
\delta(m^2_R - p^2 - 2 E_J k_{\rm in}),
\end{equation}
and similarly for the Compton channel. The fact that in the $R\to 0$ limit
$S_{q\bar{q}}$ depends only on the parton that initiates the jet can be seen
explicitly by taking the limit in Eq.~(\ref{eq:soft_aux}).  In this limit,
 only the terms proportional to $\ln r$ survive, each of which is proportional
 the jet's Casimir -- $C_F$ for quark jet and $C_A$ for a gluon jet.  Alternatively, it
can be understood by the rescaling argument from \cite{Kelley:2011aa}. To
be specific, by rescaling $R$ to $\pi/2$, the cone of the $\hat{n}_J$-jet
becomes a hemisphere. At the same time, all other Wilson lines, except
the one along the jet direction, are forced to point in the direction of
$-\hat{n}_J$. Therefore, a soft dipole contribution from partons $i$ and $J$,
where $i$ denotes an initial state parton, is identical to the hemisphere
soft function~\cite{Kelley:2011aa}, with the color factor replaced by
$-T_i\cdot T_J$ in the color generator notation~\cite{Catani:1996jh}, while
dipoles contributes from two initial state partons $i$ and $j$ vanish.
Since $(-T_1\cdot T_J-T_2\cdot T_J)=T^2_J$, summation over different
soft dipole gives a result which depends only on the jet direction and
its color content. This explains why the jet function constructed in
Eq.~(\ref{eq:jetsmallR}) depends only on the parton that initiates the
jet. Indeed, at $\mathcal{O}(\alpha_s)$, for quark and gluon jet, we have
\begin{eqnarray}
  J^{SCET}_q (m^2_R,\mu) &=& \frac{\alpha_s(\mu) C_F}{\pi m^2_R}
  \left(
    \log\frac{4E^2_J \sin^2\frac{R}{2}}{p^2_T}
    - \frac{3}{4}
  \right),
  \nn
  J^{SCET}_g (m^2_R,\mu) &=& \frac{\alpha_s(\mu)}{\pi m^2_R}
  \left(
         C_A\log\frac{4E^2_J \sin^2\frac{R}{2}}{p^2_T}
         - \frac{11}{12}C_A
         + \frac{1}{3}n_f T_F
  \right),
  \label{eq:jetsmallR2}
\end{eqnarray}
respectively, where $n_f$ is the number of light flavor and $T_F=1/2$ in QCD.
Eq.~(\ref{eq:jetsmallR2}) agrees with the corresponding $\mathcal{O}(\alpha_s)$
jet function in Eqs.~(A3) and (A4) of Refs.~\cite{Li:2012bw}, after power
expanding in jet mass, except the term proportional to $n_f T_F$, which is
missing there. From the discussion above, we see that the results presented in
Refs.~\cite{Li:2011hy,Li:2012bw} are not complete at NLL level, missing the
contribution from large angle soft cross talk, as well as the term proportional
to $n_f T_F$ in the gluon jet function. These extra global NLL contributions
are taken into account in a more recent work \cite{Dasgupta:2012hg}, which
also includes a numerical exponentiation of part of the non-global structure.

\section{Scale Choices}
\label{sec:scales}

\begin{figure}[t]
    \begin{center}
    \includegraphics[scale=0.6]{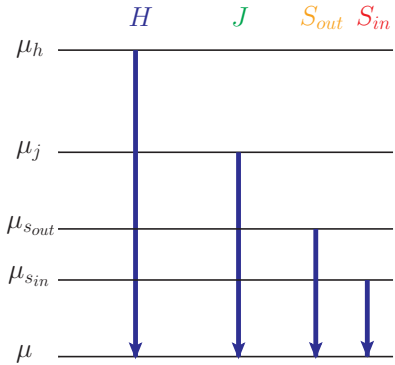}
    \caption{\label{fig:scale1}
    The renormalization group evolution of the hard, jet and soft functions, as
    shown schematically. Each function is calculated at fixed order at its natural
    scale, and they are evolved to a common scale $\mu$. Two scales emerge in the
    soft sector which results in the refactorization of the soft function.}
    \end{center}
\end{figure}

While the resummed result is formally independent of the scales $\mu_h$,
$\mu_j$, $\mu_{\in}$ and $\mu_{\out}$ as well as the factorization scale we call $\mu$,
there is residual higher-order dependence on these
scales if the perturbative expansions of the hard, jet and soft functions are
truncated at a finite order. To get a well behaved expansion with minimum
hard, jet and soft scale dependence, we want to evaluate each contribution at
its natural scale where each function does not involve large perturbative logarithms.
We evolve the hard, jet and soft functions from their natural scales to the
 factorization scale $\mu$ as shown in \fig{scale1}.

In the resummed distribution, the following ratios of scales appear
\be
    \frac{p_T^2}{\mu_h^2}\;,
    \qquad \frac{m_R^2}{\mu_j^2}\;,
    \qquad \frac{\mu_j^2}{p_T\mu_{\in}}\;,
    \qquad \frac{m_X^2-m_R^2}{p_T\mu_{\out}} \;.
\ee
Unlike the fully inclusive case
studied in \cite{Becher:2009th}, it would be impossible to minimize logarithms
of these ratios with only 3 scales $\mu_h$, $\mu_j$ and $\mu_s = \mu_{\in} =
\mu_{\out}$. Instead, we will allow $\mu_{\in}$ and $\mu_{\out}$ to be independent,
as explained in \sec{soft_fact}. When we choose a single
soft scale, the resummed SCET distribution is hopelessly different from the
{\sc pythia} output -- this is shown in \fig{500a4}.

\begin{figure}[t]
    \begin{center}
    \includegraphics[scale=1.2]{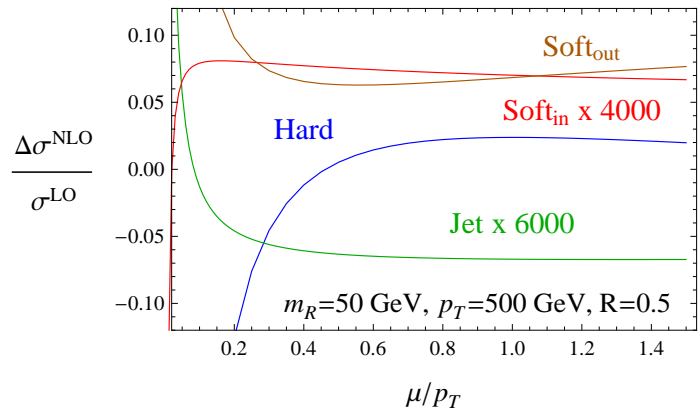}
    \caption{
      \label{fig:natural}
      The NLO corrections of the hard, jet and soft functions to the jet mass
      distribution as a function of the renormalization scale for $m_R=50$
      GeV and $p_T=500$ GeV . The natural scale of each mode is indicated by the
      of the extremum of the appropriate curve.
      The hard and $\rm soft_\out$ scales do not depend on the mass of the
      jet; they are determined by varying the scales with the NLO delta
      function contributions. The jet and $\rm soft_\in$ scales are determined
      by varying the scales with the NLO jet mass distribution. The jet and
      $\rm soft_\in$ curves are scaled up to plot with the
      ones for the hard and $\rm soft_\out$.
    }
    \end{center}
\end{figure}
\begin{figure}[t]
    \begin{center}
    \includegraphics[scale=1]{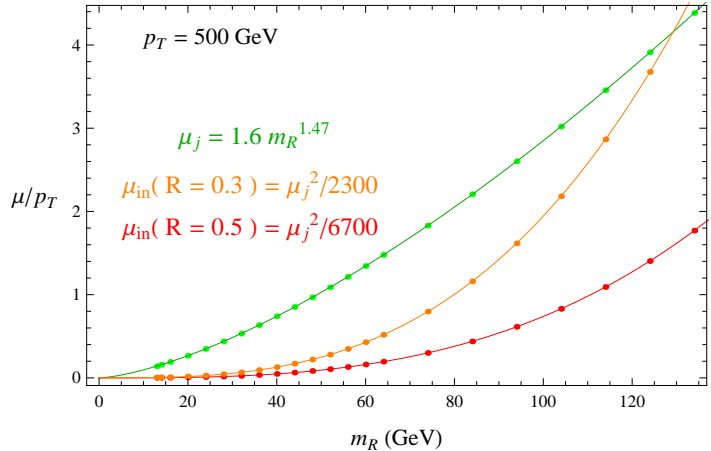}
    \caption
    {
      \label{fig:mujs}
      The jet and $\rm soft_\in$ scales as a function of $m_R$ for $p_T=500$
      GeV with two different cone sizes. Dots are extrema and curves are our fits.
 Note that $\mu_{\in}$ is cone
      size dependent and increases when the cone gets smaller. However,
      $\mu_j^2/\mu_{\in}$ is $m_R$ independent, which is a manifestation
      of the factorization theorem. The numerically determined jet and
      $\rm soft_\in$ scales fit nicely with power law curves ($m_R$ in
      GeV): $\mu_j=1.6~m_R^{1.47}$, $\mu_{\in}(R=0.3)=\mu_j^2/2300$, $\mu_{\in}(R=0.5)=\mu_j^2/6700$, and they are considerably
      higher than the naive choices of $\mu_j=m_R$ and $\mu_{\in}=m_R^2/p_T$.
   }
  \end{center}
\end{figure}

The naive scale choices are
\be
    \mu_h=p_T\;,
    \qquad \mu_j=m_R\;,
    \qquad \mu_{\in}=\frac{m_R^2}{p_T}\;,
    \qquad \mu_{\out}=\frac{m_X^2-m_R^2}{p_T} \;,
\ee
These would eliminate the large logarithms at each energy scale at the partonic
level. However, when the partonic cross section is integrated against the
PDFs, $\mu_{\out}$ will get arbitrarily close to zero and hit the Landau
pole singularity of the running coupling. The problem is that the natural
$\mu_\out$ scale choice depends on an unphysical quantity, namely $m_X^2 -
m_R^2$. We can avoid the Landau pole and minimize the effects of the large
logarithms by instead choosing the scales numerically to depend only on
observables~\cite{Becher:2007ty}. In fact, it is not necessary to eliminate the
logarithms in the unphysical, partonic cross section, but rather we want to
eliminate the large logarithm in the physical, hadronic cross section.

To determine the natural RG scales, we follow the approach used
in~\cite{Becher:2007ty,Becher:2009th,Becher:2011fc,Becher:2012xr}. We include
separately the NLO corrections for the hard, jet or soft functions and vary
the relevant scale to find out which scale minimizes the variation of the
NLO distribution. For example, \fig{natural} shows these NLO corrections for the 8 TeV LHC
with a $p_T= 500$ GeV photon and
$m_R=50$ GeV. The extrema of these curves indicate a natural scale for each mode,
and one can see that there is a natural hierarchy of
the various matching scales.
The jet and $\rm soft_\in$ scales are jet mass dependent, whereas the hard and $\rm
soft_\out$ scales are jet mass independent (their NLO contributions have no dependence on $m_R$).
We fit the jet mass dependence by power law curves, examples of which are shown in \fig{mujs} for different
jet sizes. For example, with $R=0.5$, the scale choice we extract are
\be
    \mu_h=p_T;,
    \qquad \mu_j=1.6~m_R^{1.47}\;,
    \qquad \mu_{\in}=\frac{\mu_j^2}{6700}\;,
    \qquad \mu_{\out}=280~{\rm GeV}  .
\ee
We choose the factorization scale to match the hard scale, $\mu = \mu_h$ (a high factorization scale is not natural
in SCET, however it corresponds most closely to the scales where the PDFs have been fit).
The renormalization group is then used to run between these scales to producing the resummed  jet mass distribution.
 For simplicity, we extract these
curves at fixed $p_T$ and $y$ for the photon, although one could easily repeat
this exercise integrated over some window of photon kinematics.

\section{Results}
\label{sec:results}

\begin{figure}
    \begin{center}
    \includegraphics[scale=0.8]{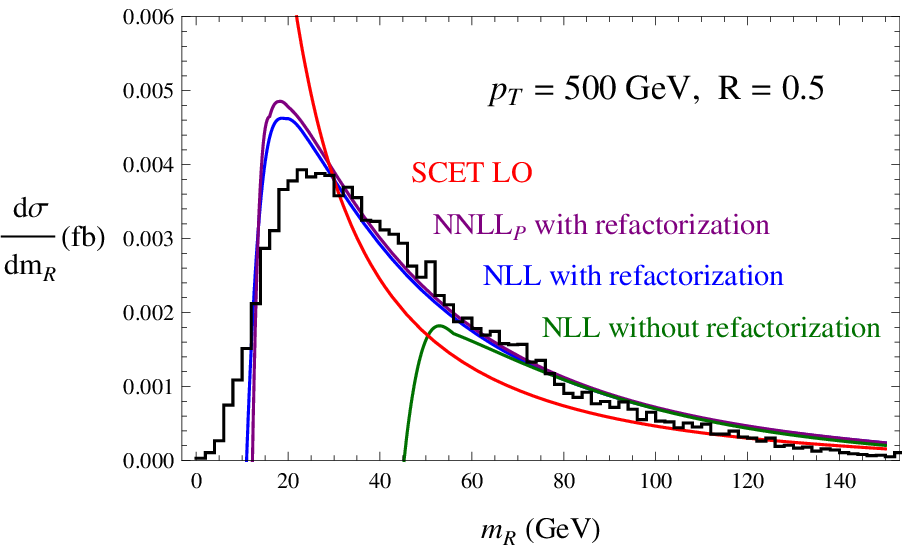}
    \includegraphics[scale=0.8]{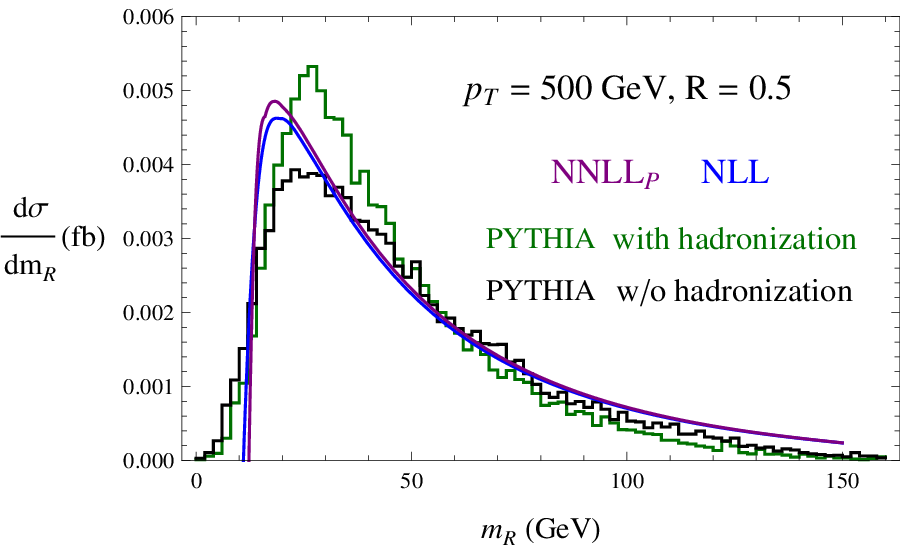}
    \caption{
      \label{fig:500a4}
      Comparison of the jet mass distributions with
      different orders of precision (curves) for $p_T=500$ GeV and $R=0.5$ to {\sc pythia} (histograms).       
      The right plot shows the effects of hadronization in {\sc pythia} as
      compared to resummed distributions.
      }
    \end{center}
\end{figure}

To test our predictions, we compare our results to the output from {\sc pythia
8}~\cite{Sjostrand:2007gs} for the LHC at $\ecm=8$ TeV. We use the MSTW 2008
NLO PDFs~\cite{Martin:2009iq} both for event generation and theory calculation.
The event generation is made with the hadronization turned off, except for one curve
on the right hand side of \fig{500a4}, where you can see that it affects the peak region.
For the rest of plots, we keep the initial state and final state radiation,
but we turn off hadronization, underlying event and multiple interactions.
These additional effects are important, but we postpone their consideration to
future work. The specific observable we use is outlined in \ssec{observable},
and we test several values of $R$. The jet mass distribution is calculated and
compared to {\sc pythia} in a small window of the photon transverse momentum
and rapidity $(\Delta p_T,\Delta y)=(10~{\rm GeV},0.1)$, centered around $p_T =
500$ or $2000$ GeV and $y =0$, for various sizes of $R$.

\begin{figure}[t]
    \begin{center}
    \includegraphics[scale=0.8]{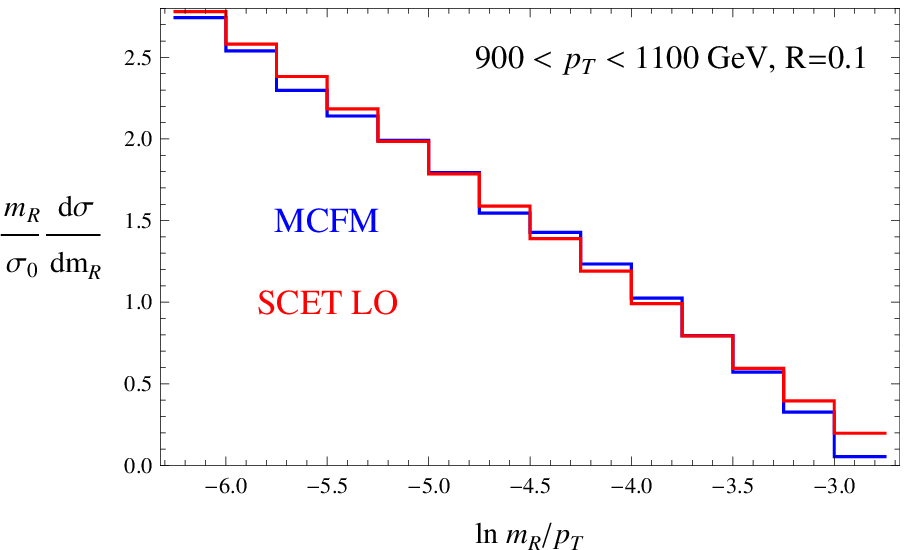}
    \includegraphics[scale=0.87]{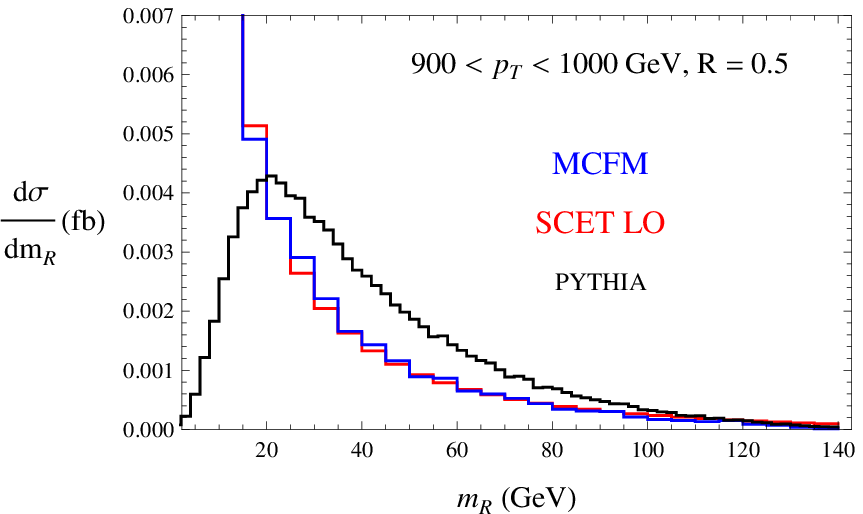}
    \caption{The left panel shows a comparison of the leading order jet mass distirbution calculated with MCFM
and the prediction from expanding the resummed result to leading order (SCET LO). This demonstrates
that SCET reproduces the singularities of QCD at leading order. 
The right panel compares MCFM and SCET to {\sc pythia}, with obvious disagreement, indicating the importance of 
resummation.
      \label{fig:mcfm}
      }
    \end{center}
\end{figure}

Since SCET contains all the physics in the small $m_R$ regime, it should
reproduce the singular structure of full QCD at leading order. We check this
by comparing the fixed order expansion of SCET to the exact leading order
distribution, using MCFM~\cite{Campbell:2010ff}, in the far singular region.
The result is shown in \fig{mcfm}, which shows very good agreement down to very
small $m_R$. We also compare MCFM with {\sc pythia}, and there is no region of
$m_R$ in which the QCD NLO calculation agrees, which suggests the necessity of
resummation.

We calculate the jet mass distribution at full next-to-leading log (NLL)
precision, which involves the two-loop cusp anomalous dimension and the one
loop anomalous dimensions of the hard, jet and soft functions. In fact, all
ingredients are known for NNLL resummation, except for the contribution
of non-global logarithms. We estimate the effect of these missing NGLs in
\sec{NGL}. We denote the partial next-to-next-to-leading log resummed result
as $\rm NNLL_p$, which has the three loop cusp anomalous dimension and the
two-loop anomalous dimensions of the hard, jet and full soft functions. We
have only included those parts of the $\mathcal{O}(\alpha_s^2)$ soft anomalous
dimension that can be extracted from the RG invariance of the cross section,
which does not include the angle-dependent pieces.

\begin{figure}
    \begin{center}
    \includegraphics[scale=0.81]{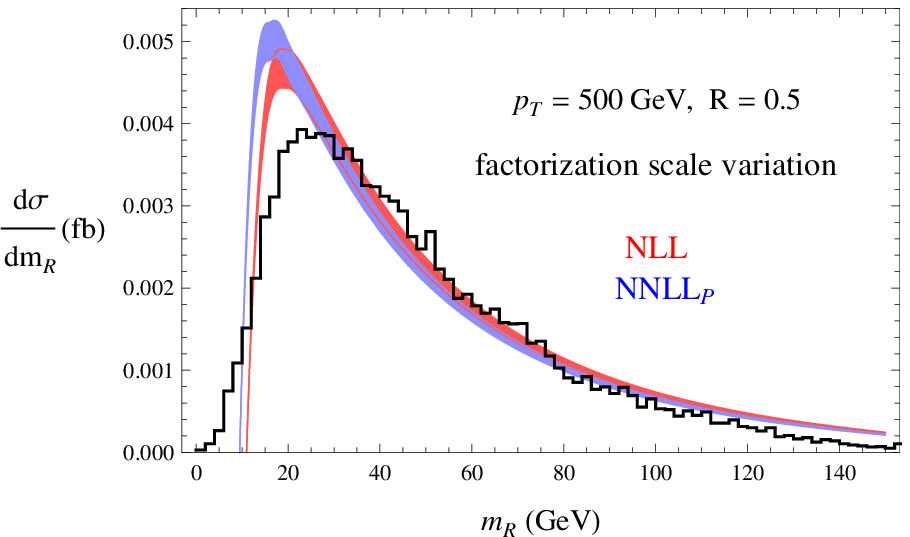}
    \includegraphics[scale=0.81]{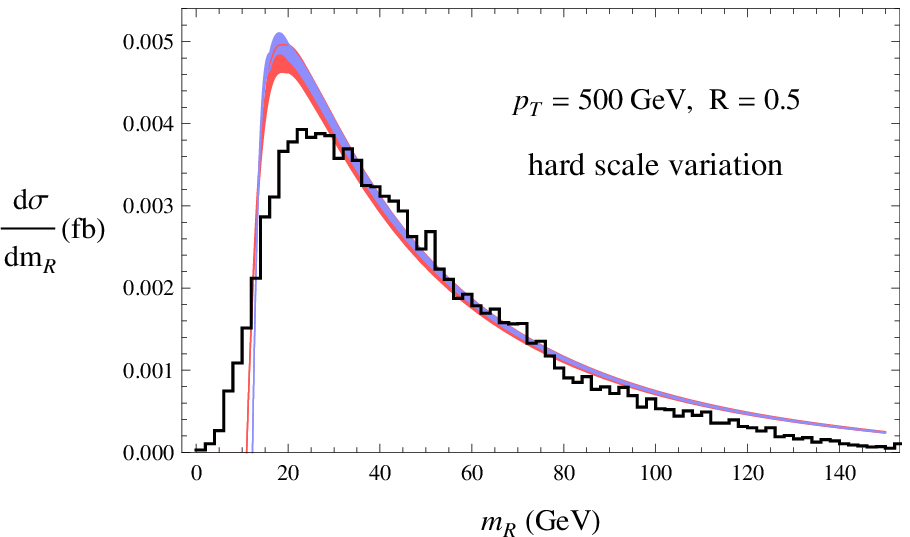}
    \includegraphics[scale=0.81]{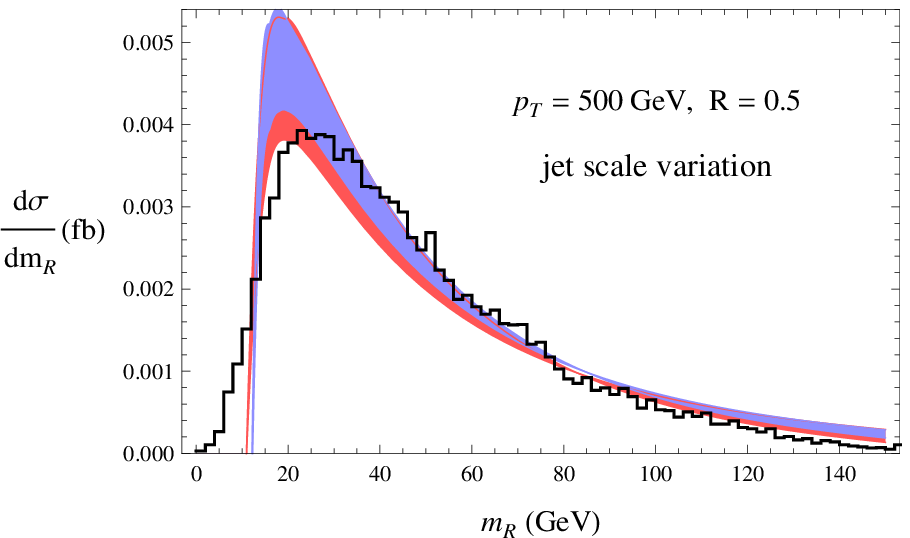}
    \includegraphics[scale=0.81]{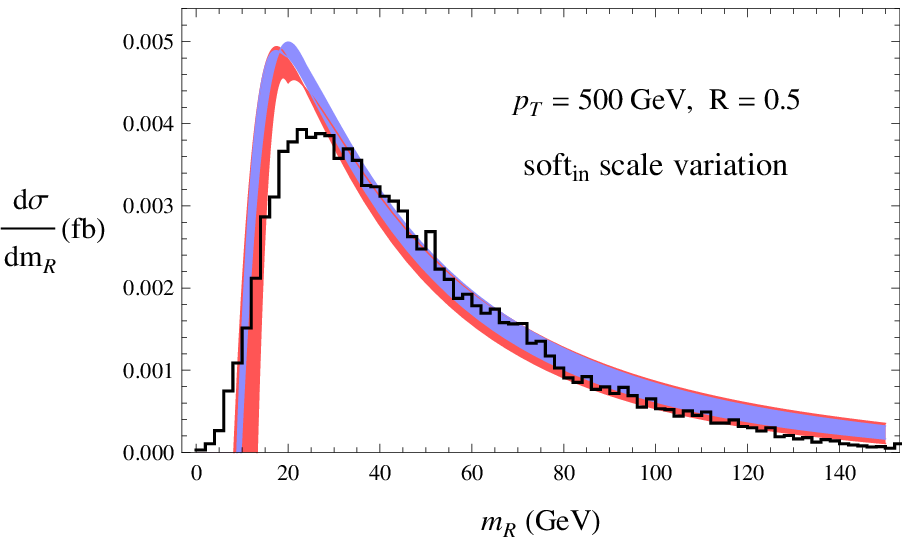}
    \includegraphics[scale=0.81]{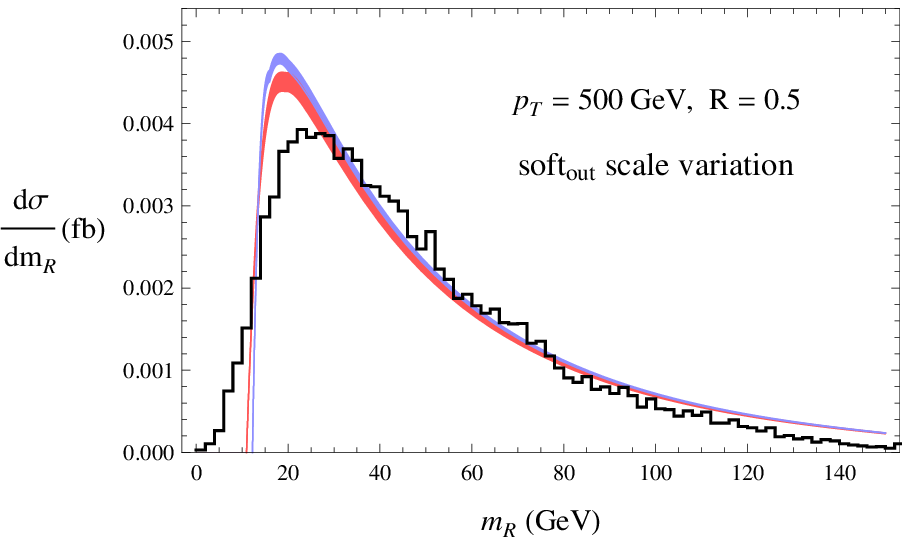}
    \caption{\label{fig:500hj}
    Scale uncertainties for $p_T=500$ GeV and $R=0.5$. The red bands are NLL and the blue bands are $\rm NNLL_p$. The {\sc pythia} result is the histogram shown in black.}
    \end{center}
\end{figure}

\begin{figure}
    \begin{center}
    \includegraphics[scale=0.81]{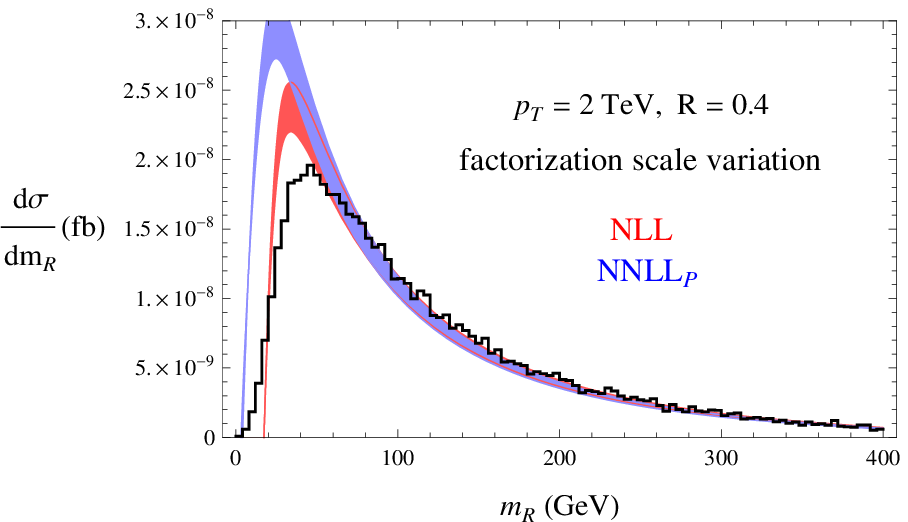}
    \includegraphics[scale=0.81]{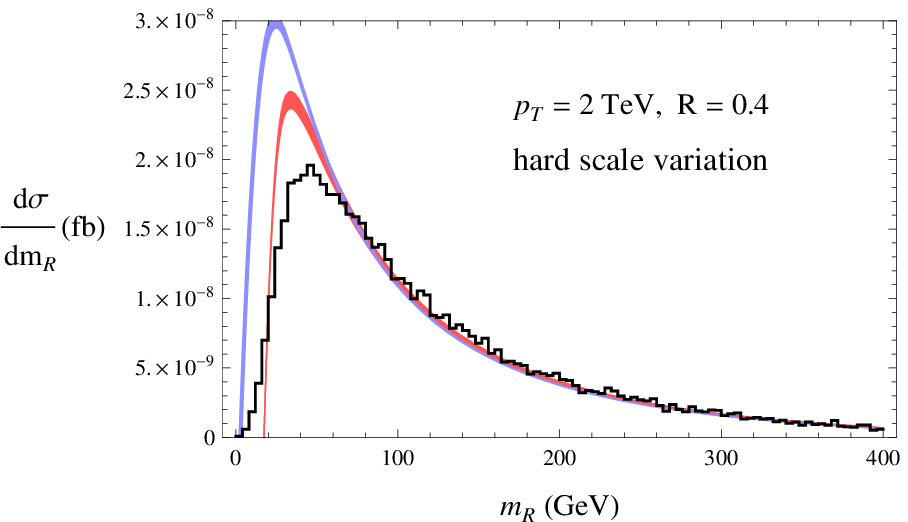}
    \includegraphics[scale=0.81]{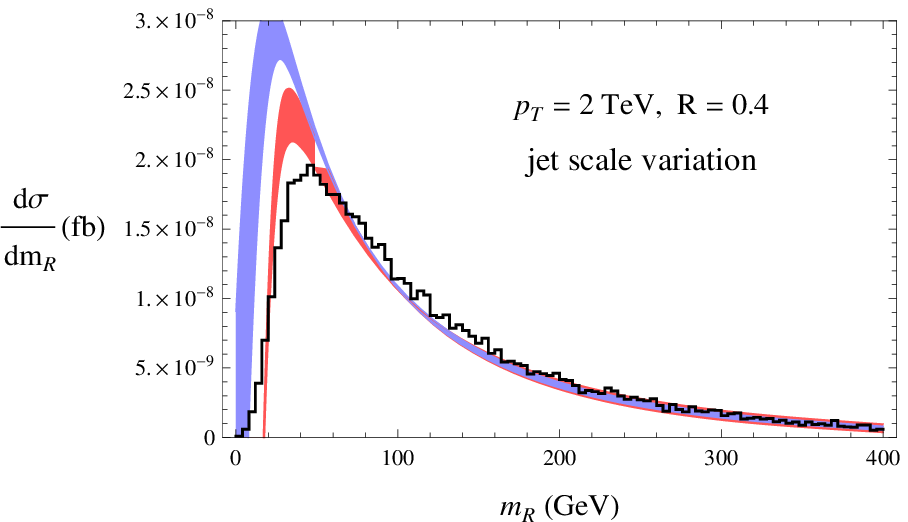}
    \includegraphics[scale=0.81]{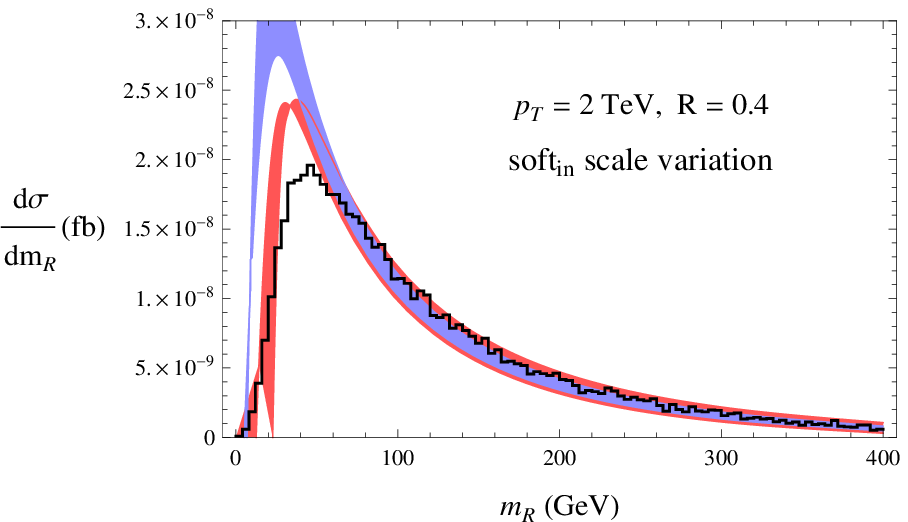}
    \includegraphics[scale=0.81]{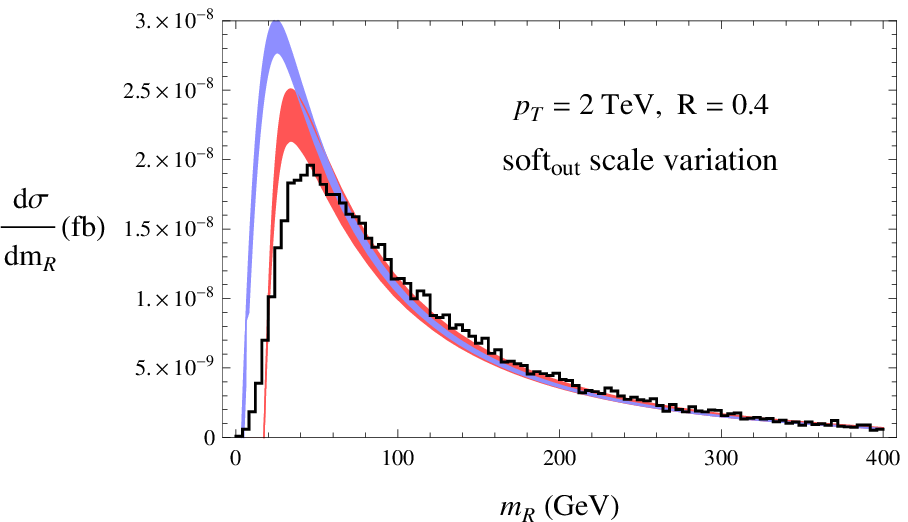}
    \caption{\label{fig:2000hj} Scale uncertainties for $p_T=2$ TeV and $R=0.4$.}
    \end{center}
\end{figure}

\begin{figure}
    \begin{center}
    \includegraphics[scale=0.78]{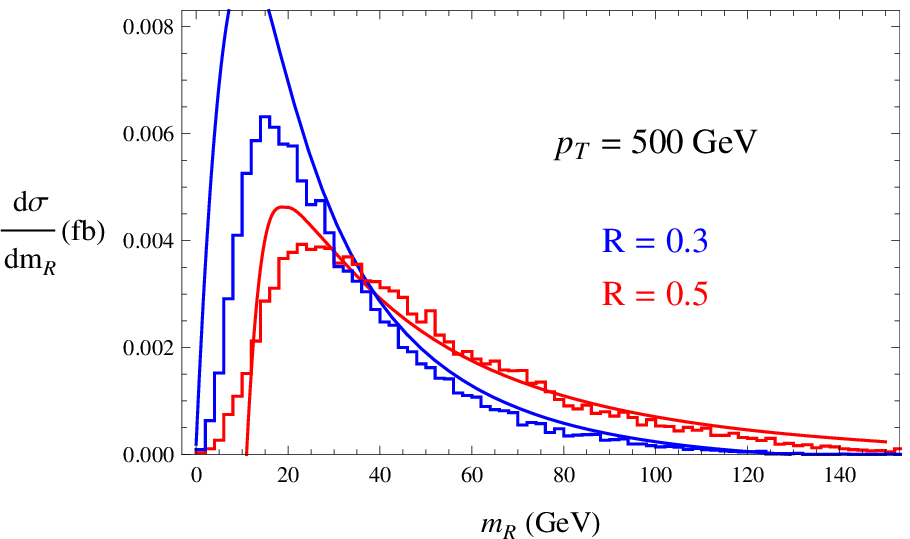}
    \includegraphics[scale=0.85]{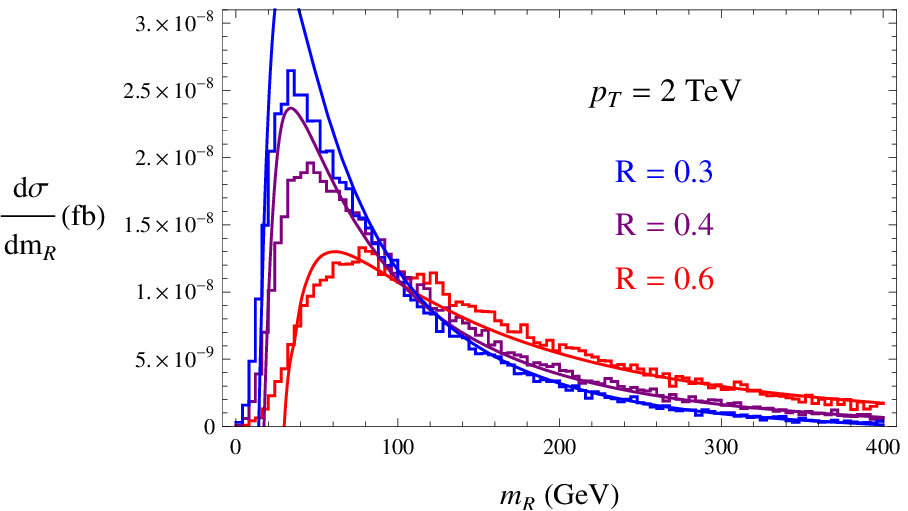}
    \caption{\label{fig:resum} The resummed NLL jet mass distribution for different cone sizes for  $p_T=500$ GeV and 2 TeV (curves)
compared to {\sc pythia} (histograms).}
    \end{center}
\end{figure}

Figure \ref{fig:500a4} shows the jet mass distributions for the $p_T=500$ GeV
events at different orders of precision.
For this $p_T$, we compute the
jet mass distribution with a maximum cone size of $R=0.5$ to avoid the jet being contaminated by the beam. We normalize the SCET distributions with the NLO
QCD cross section as determined by {\sc pythia} 8, scaled by the QCD NLO $K$-factor
for the cross-section with the jet rapidity restriction ($K\sim0.8$ for a 500 GeV photon), which we compute with our own code.
We do not match to the exact LO QCD calculation which would account the power corrections in the
tail. In the peak region, where most of the events lie, these power corrections are small.
The scale uncertainties for the
resummed result include variation of the factorization scale $\mu_f=\mu$, the hard scale $\mu_h$, the jet scale
$\mu_j$, and the soft scales $\mu_{{\rm in}}$ and $\mu_{{\rm out}}$,
Figure \ref{fig:500hj} shows the uncertainty bands for separate variation of the scales
between $\frac{1}{2}\mu_i<\mu<2\mu_i$ for $i = f, h, j, s_\in, s_\out$,
for $p_T=500$ GeV and $R=0.5$ jets.
Additional comparisons for  $p_T = 2$ TeV and $R=0.4$ are shown in Figures \ref{fig:2000hj}
and \ref{fig:resum}. The higher the transverse momentum of the photon, the closer to threshold, so we expect that threshold resummation will be more effective in this case.

\section{The Role of Non-Global Logarithms}
\label{sec:NGL}

As mentioned in Section \ref{sec:soft_fact}, although we were able to refactorize the
soft function into a soft-collinear part, whose natural scale is associated
with the soft modes within the jet, the remainder soft function still
depended on multiple scales $S_r=S_r(k_\in,k_\out,\mu)$. Thus we cannot
guarantee that all large logarithms in the jet mass distribution are resummed.
The residual dependence of the remainder soft function on two scales is the
problem of non-global structure. In the absence of a complete understanding of
non-global structure, and how the non-global logs (NGLs) might be resummed,
we will content ourselves with an estimate of how non-global structure might
affect the jet mass distribution. We start by drawing on the lessons learned
when considering $e^+e^- \to $ dijets, where there have been several studies
\cite{Banfi:2010pa,Kelley:2011ng,Hornig:2011iu,KhelifaKerfa:2011zu,
Kelley:2011aa}.

A two-loop calculation of the soft function for hemispherical jets was
performed in \cite{Kelley:2011ng,Hornig:2011iu} and for cone or anti-$k_T$
jets, with an out-of-jet veto, in \cite{Kelley:2011aa}. An intriguing
observation from these studies is that in both examples, the NGLs arose from
combinations of various logarithms of a single scale, coming from integrals in
separate phase space regions. To give a specific example, in the calculation of
the cumulative doubly differential dijet mass distribution for hemispherical
jets, the details of which can be found in \cite{Kelley:2011ng}, it was shown
that the leading non-global logarithm is of the form
\be
 - \frac{4\pi^2}{3} C_F C_A \ln^2\frac{E_L}{E_R} \;.
\ee
This is the coefficient of $\left(\frac{\alpha_s}{4\pi}\right)^2$ in the soft
function, with $E_L$ and $E_R$ cumulant variables corresponding to integrals
over $k_{L} = \nbar \cdot P_{X_L}$ and $k_{R} = n \cdot P_{X_R}$, respectively,
where $P_{X_{L}}$ and $P_{X_{R}}$ are the total momentum of the soft radiation
propagating in the left or right hemispheres and $n = (1, \hat{n})$ denotes the
thrust axis. This double logarithm came from the sum of three contributions
from the two gluon real emission graphs off the two Wilson lines
\be
\label{eq:NGLs}
\begin{tabular}{lcl}
$\ds - \frac{4\pi^2}{3}  C_F C_A \ln^2 \frac{E_L}{\mu}$       \label{first}
&& \text{both gluons in left hemisphere} \nn[8pt]
$\ds - \frac{4\pi^2}{3}  C_F C_A \ln^2 \frac{E_R}{\mu}$
&& \text{both gluon in right hemisphere} \nn[8pt]
$\ds + \frac{8\pi^2}{3} C_F C_A \ln\frac{E_L}{\mu} \ln\frac{E_R}{\mu}$
&& \text{one gluon in each hemisphere}
\end{tabular}\;.
\ee
One therefore expects a similar sum of phase space regions to produce the
leading non-global logarithm for the soft function we need in this paper.

Now, we have already refactorized the soft function as
\be
  S(k_\in,k_\out,\mu)=\bS(k_\in,\mu)\rS(k_\in,k_\out,\mu) \;,
\ee
with the regional
soft function $\bS(k_\in,\mu)$ defined to have all the radiation going into the
jet. Thus we expect a double logarithm similar to the first term in \eqn{NGLs}
to contribute at two loops
%
%
\be
\label{eq:extra_log}
\bS(\kappa_\in,\mu)=\cdots + \left( \frac{\alpha_s}{4\pi} \right)^2 \GNGL \ln^2 \frac{\kappa_\in}{\mu} \;,
\ee
where $\GNGL$ might be $-\frac{4\pi^2}{3} C_F C_A$, but is yet unknown.
Since this term has $\mu$-dependence, it will contribute to the anomalous
dimension of $\bS$. Moreover, it will contribute a term with $\ln \mu$ in the
anomalous dimension. This is unusual since, for a global observable, one
normally expects that all of the $\ln \mu$ dependence in the anomalous dimension is
proportional to $\gamma_{\rm cusp}$. We therefore expect similar non-cusp $\ln \mu$
dependence in the anomalous dimension of the regional soft function for the jet mass distribution as well.

With this educated guess, we expect that the two-loop regional soft function anomalous
dimension for the annihilation channel has the form
\be
\label{eq:gamma2_aux}
\gamma_{\bS}
=
\left(\frac{\alpha_s}{4\pi}\right)^2 \left[
\ds
  \left( 2C_A \Gamma_1  + \Gamma^{\rm NGL} \right)
  \ln \left( \frac{1+\beta^2}{\beta} \frac{\kappa_\in}{\mu} \right)  +C_A\gamma^s+\gsreg
\right]
\ee
and similarly for the Compton channel, with $C_F \to C_A$ by Casimir scaling.
Here, $\Gamma_1$ is the two-loop cusp anomalous dimension:
\bea
\label{eq:Gamma_1}
\Gamma_1 = 4 \left(  \frac{67}{9}  - \frac{\pi^2}{3}\right) C_A-\frac{20}{9}n_f T_F \;,
\eea
and $\GNGL$ and $\gsreg$ are unknown. By RG invariance, this implies that the
anomalous dimension of the residual soft function must be
\be
\label{eq:gamma2_residual}
\gamma_{\rS}
=
\left(\frac{\alpha_s}{4\pi}\right)^2 \left[
\ds
  -4C_F \Gamma_1
  \ln \left( \frac{1+\beta^2}{\beta} \frac{\kappa_\out}{\mu} \right)
  - \Gamma^{\rm NGL}
  \ln \left( \frac{1+\beta^2}{\beta} \frac{\kappa_\in}{\mu} \right)
  -2C_F \gamma^s-\gsreg
\right]
\ee
where $\gamma_s$ is the two-loop direct-photon soft function anomalous dimension from\cite{Becher:2009th}:
\bea
\label{eq:gamma_s}
\gamma^s =
\left(
  28\zeta_3
  -\frac{808}{27}
  + \frac{11\pi^2}{3}
\right) C_A
+
\left(
\frac{224}{27}
-\frac{4\pi^2}{9}
\right)
n_f T_F \;.
\eea
The terms proportional to $\Gamma_{1}$ and $\gamma_s$ in \eqn{gamma2_aux} and
\eqn{gamma2_residual}, and also a known term proportional to the 3-loop cusp
anomalous dimension, account for the global logarithms at NNLL. The $\GNGL$
term parameterizing the unknown leading NGL. The $\gsreg$ anomalous dimension
accounts for sub-leading NGLs, and the remaining non-global structure in
the finite part of the two-loop soft function has effects which are formally
N${}^3$LL.

\begin{figure}[t]
   \begin{center}
   \includegraphics[scale=0.81]{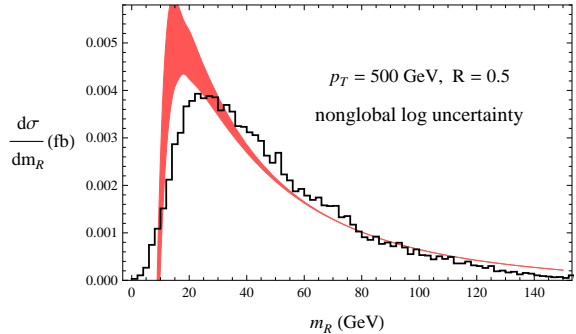}
   \caption{\label{fig:500ngl} Estimation of the effect of leading
non-global log resummation. Note that it affects significantly only the peak region of the distribution.}
   \end{center}
\end{figure}

Without performing a two-loop calculation, the expressions for $\GNGL$ and
$\gsreg$ are unknown. We estimate the effect of the missing terms by varying
$\GNGL$ between $\pm 100$, since $100\sim 4 C_F \Gamma_1 \sim \frac{8\pi^2}{3}
C_F C_A$. The result of performing such a variation is shown in \fig{500ngl}.
The band in this plot shows a reasonable expectation of the improvement one
could expect if the leading non-global logarithm could be resummed. We see that
the NGL only affects significantly the distribution in the peak region.

%

\section{Conclusions}
\label{sec:conc}

In this paper we have calculated the distribution of jet mass at a hadron
collider, for events in which the jet recoils against a hard photon. Our
calculation includes resummation at the next-to-leading logarithmic level (NLL)
which involves both final state radiation of the jet and initial state
radiation of the colliding partons. We also have included resummation of
all the global logarithmic terms at next-to-next-to-leading logarithmic level
producing a distribution we label NNLL${}_p$ (``p'' for partial, since the
non-global logarithmic terms have not been resummed). Our approach is based on
expanding around the threshold limit, where the photon has very large momentum.
By demanding the photon be hard, we force the hadronic final state to be that
of a single jet for which a simple factorization formula exists. Our result is
differential in the jet and photon rapidities and transverse momenta, although
we assume the photon and jet have equal and opposite transverse momenta, which
is true at leading power.

We have compared our theoretical calculation to the output from {\sc pythia},
and find very good agreement. Although {\sc pythia} is only formally accurate
to the leading-logarithmic level, it has elements of subleading logarithms,
and is expected to be in good agreement with collider data. Since we are able
to include color coherence effects, from initial state radiation, as well as
subleading logarithms, our calculation is more precise than {\sc pythia},
as far as perturbative QCD is concerned. Since {\sc pythia} also includes
hadronization it is more likely to describe the data well for very small
jet mass. It would be interesting to compare both directly to collider data from
the LHC when it becomes available.

The two scales in our distribution, the photon $p_T$ and the jet mass $m_R$
lead to non-global logarithms. This is because in our approach there are
two thresholds, $m_R \to 0$ and $m_X \to 0$, where $m_X$ is the mass of
everything-but-the-photon. In particular, there are two soft scales, $k_\in$
associated with soft radiation in the jet, related to $m_R$, and $k_\out$
associated with radiation out of the jet, which can be related to $m_X^2 -
m_R^2$. Non-global logarithms of the ratio of these scales make resummation
difficult beyond the NLL level. An alternative calculation would be to impose
a jet veto at a scale similar to the soft scale. Then the non-global structure
would reduce to a single number, which one might estimate or argue to be small.
However, for an inclusive jet mass calculation, it seems impossible to avoid
non-global structure and therefore resumming non-global logarithms is necessary
for NNLL resummation.

Although we have not been able to resum the non-global logs in this paper, we
found that if one ignores them completely, by choosing a single scale for the
entire soft function, the distribution is completely wrong. This is because,
using SCET alone, the two soft scales are not distinguished. Instead, we
observe that contributions to the soft function coming from radiation going
entirely into the jet can be consistently factorizes off. There are many ways to
refactorize the soft function, and we make one particular choice, preserving
the color structure associated with the hard directions. We give an operator
definition to this regional soft function and calculate it at one loop. The
regional soft function has a natural scale associated with soft/$n_J$-collinear
modes. Allowing ourselves to pick two different soft scales, which we do
numerically, gives a result which is in very good agreement with {\sc pythia},
in the region that our calculation can be trusted.

There are many directions in which this work can be continued. It would be
interesting to calculate the regional soft function at two loops, to see its
non-global structure. One might hope that, since we expect an anomalous
dimension for this soft function to have $\mu$ dependence related to the
leading non-global logarithm, that with further insight non-global logarithms
could be resummed. Then one could produce an NNLL calculation of jet mass.
Other related applications would include a calculation of the jet mass in
dijet events, for which there is already data~\cite{ATLAS-CONF-2011-073} and the one-loop
hard function and anomalous dimensions have already been prepared~\cite{Kelley:2010fn}, or
 a calculation
of jet mass in direct photon events including a jet veto. With a jet veto one
can force single-jet kinematics well away from threshold and the size of the
non-global structure can be controlled. However, the calculation would be significantly
more complicated than what we have done here.
 In addition, it would be interesting to
pursue the consideration of the jet mass distribution using different jet
algorithms.


\section*{Acknowledgments}
YTC would like to thank Hsiang-nan Li for discussion.
YTC, RK and MDS were supported in part by the Department of Energy, under grant
DE-SC003916. HXZ was supported by the National Natural Science Foundation of
China under grants No.~11021092 and No.~10975004.

\bibliographystyle{JHEP3notitle}
\bibliography{direct_photon}

\end{document}